\documentclass[11pt,twoside]{article}
\usepackage{graphicx,amssymb,epsfig,epsf}
\usepackage{times}
\usepackage{a4wide}
\usepackage{cite}
\usepackage{slashed}
 \usepackage{fancyhdr}
 \advance \headheight by 3.0truept       

\newlength{\nseparation}
\newenvironment{nfigure}[1]
        {\begin{figure}[#1]\hrule\vspace{\nseparation}\par}
        {\vspace{\nseparation}\par \hrule \end{figure}}

\usepackage{color}

\newcommand{\Bbar}{\,\overline{\!B}}
\newcommand{\Kbar}{\,\overline{\!K}}
\newcommand{\bbd}{\ensuremath{B_d\!-\!\Bbar{}_d\,}}
\newcommand{\bbs}{\ensuremath{B_s\!-\!\Bbar{}_s\,}}
\newcommand{\bbq}{\ensuremath{B_q\!-\!\Bbar{}_q\,}}
\newcommand{\bb}{\ensuremath{B\!-\!\Bbar{}\,}}
\newcommand{\bbms}{\bbs\ mixing}
\newcommand{\bbmd}{\bbd\ mixing}
\newcommand{\bbmq}{\bbq\ mixing}
\newcommand{\bbm}{\bb\ mixing}
\newcommand{\kk}{\ensuremath{K\!-\!\Kbar\,}}
\newcommand{\kkm}{\kk\ mixing}

\newcommand{\dm}{\ensuremath{\Delta M}}

\newcommand{\eq}[1]{Eq.~(\ref{#1})}

\newcommand{\ps}{p\hspace{-0.4em}/\hspace{0.02em}}
\newcommand{\fig}[1]{Fig.~\ref{#1}}
\newcommand{\bra}[1]{\ensuremath{\langle #1 |}}
\newcommand{\ket}[1]{\ensuremath{| #1 \rangle }}

\begin{document}
\thispagestyle{plain}
\parbox[t]{0.4\textwidth}{TTP09-28}
\hfill August 2009 \\

\begin{center}
\boldmath
{\Large \bf Chirally enhanced corrections to 
 FCNC processes in the generic MSSM}\\
\unboldmath
\vspace*{1cm}
\renewcommand{\thefootnote}{\fnsymbol{footnote}}
Andreas Crivellin and Ulrich Nierste \\
\vspace{10pt}
{\small
    {\em Institut f\"ur Theoretische Teilchenphysik\\
               Karlsruhe Institute of Technology,
               Universit\"at Karlsruhe, \\76128 Karlsruhe, Germany}} \\
\normalsize
\end{center}

\begin{abstract}
  Chirally enhanced supersymmetric QCD corrections to FCNC processes are
  investigated in the framework of the MSSM with generic sources of flavor
  violation. These corrections arise from flavor-changing self-energy diagrams
  and can be absorbed into a finite renormalization of the squark-quark-gluino
  vertex. In this way enhanced two-loop and even three-loop diagrams can be
  efficiently included into a leading-order (LO) calculation.  Our corrections
  substantially change the values of the parameters $\delta^{d\,LL}_{23}$,
  $\delta^{d\,LR}_{23}$, $\delta^{d\,RL}_{23}$, and $\delta^{d\,RR}_{23}$
  extracted from $\rm{Br}[B\to\rm{X_s}\gamma]$ if $\tan\beta$ is large.  We
  find stronger (weaker) constraints compared to the LO result for negative
  (positive) values of $\mu$. The constraints on $\delta^{d\,LR,RL}_{13}$ and
  $\delta^{d\,LR,RL}_{23}$ from \bbd\ and \bbms\ change drastically if the
  third-generation squark masses differ from those of the first two
  generations. \kkm\ is more strongly affected by the chirally enhanced loop
  diagrams and even sub-percent deviations from degenerate down and strange
  squark masses lead to profoundly stronger constraints on
  $\delta^{d\,LR,RL}_{12}$.
\end{abstract}

\section{Introduction}
Processes involving Flavor-Changing Neutral Currents (FCNCs) are of great
importance for supersymmetric model building, because they probe the
supersymmetry-breaking sector with an enormous sensitivity. In the Minimal
Supersymmetric Standard Model (MSSM) with generic flavor structure FCNC
processes are mediated by the strong interaction through diagrams involving
squarks and gluinos. Today's precise experimental data on flavor violation in
K, D and B physics leave little room for mechanisms of flavor violation other
than the established Cabibbo-Kobayashi-Maskawa mechanism of the Standard Model
(SM).  This incompatibility of the generic MSSM with experiment is known as
the SUSY flavor problem. It points towards a mechanism of supersymmetry
breaking which is dominantly flavor-blind and only feels the flavor violation
of the Yukawa sector.
From a phenomenological point of view one would like to assess the
possible size of the deviations from this scenario of
\emph{minimal flavor violation (MFV)}. In an appropriately chosen basis
of the (s)quark superfields (the super-CKM basis) one can parameterize
the deviations from MFV by flavor-off-diagonal entries of the squark
mass matrices. By confronting the squark-gluino mediated amplitudes 
with precision data one can constrain the sizes of these off-diagonal 
squark-mass terms. Once superparticles are discovered and their masses
are determined, these FCNC analyses may lead to insights into the
mechanism of supersymmetry breaking. 

{When the MSSM with explicit soft breaking was first
written down in \cite{Dimopoulos:1981zb} it was already realized, that
these breaking-terms are potential sources of flavour
violation. Elaborating further on this topic, Ref. \cite{Ellis:1981ts,
Barbieri:1981gn,Duncan:1983iq,Donoghue:1983mx} discovered that
neutralinos and gluinos have flavor-changing couplings once the quark
and squark fields are rotated into the physical basis with diagonal
mass matrices.}  In order to simplify the calculations and to
parameterize the non-minimal sources of flavor violation in a generic
way the mass insertion approximation (MIA) was invented
\cite{Hall:1985dx}.  MIA amounts to an expansion in the small
flavor-changing squark mass terms to lowest non-vanishing order.
Later the authors of Ref.~\cite{Gabbiani:1988rb,Bertolini:1990if}
studied FCNCs with non-minimal sources of flavor violation in the
context of various supersymmetric grand unified theories. In
Ref.~\cite{Hagelin:1992} an explicit calculation of the gluino boxes
contributing to $\Delta F=2$ processes and a matching to the effective
Hamiltonian of flavor physics was carried out for the first time,
using both MIA and the exact diagonalization of the squark mass
matrices.  In the seminal 1996 paper of Gabbiani et.~al.\
\cite{Gabbiani:1996} FCNC data (on B mixing, K mixing, $\mu\to
e\gamma$ and $b\to s\gamma$) were used to constrain the off-diagonal
elements of the sfermion mass matrices, while the diagonal ones were
constrained from the electric dipole moments (and from a fine tuning
argument). In that work the mass insertion approximation was used and
QCD corrections were briefly discussed. After the QCD corrections to
the full $\Delta F=2$ Hamiltonian were calculated at leading order
(LO) \cite{Bagger:1997gg} and next-to-leading order (NLO)
\cite{Ciuchini:1997bw} an improved calculation of Kaon mixing in the
generic MSSM has lead to more accurate bounds on the off-diagonal
elements of the squark mass matrix
\cite{Bagger:1997gg,Ciuchini:1998ix}.  {In 1999 an
extensive study of $b\to s \gamma$ with LO Wilson coefficients
\cite{Borzumati:1999} was performed. Ref.~\cite{Besmer:2001cj}
extended the study of {Ref.~}\cite{Borzumati:1999} by
{including} the MFV effects.}
Ref.~\cite{Becirevic:2001} has addressed $\rm{B}$ mixing at NLO,
Ref.~\cite{Ciuchini:2002uv,Ciuchini:2006dx} contains an update of the
constraints from $b\to s$ transitions and Ref.~\cite{Ciuchini:2007cw}
has discussed D mixing. In the meantime also the Wilson coefficients
for $\Delta F=2$ processes have been calculated at NLO
\cite{Ciuchini:2006dw}.  {The importance of
chirally-enhanced beyond-LO corrections in the generic MSSM was
discussed in a series of articles by {Foster et al.}
\cite{Foster}.}
Recently, constraints on the off-diagonal elements of the squark mass
matrix have been obtained from a fine-tuning argument applied to the
renormalization of the Cabibbo-Kobayashi-Maskawa (CKM) matrix
\cite{Crivellin:2008mq} and from a loop-induced effective right-handed
W coupling \cite{Crivellin:2009sd}.

\begin{nfigure}{t}
\includegraphics[width=0.9\textwidth]{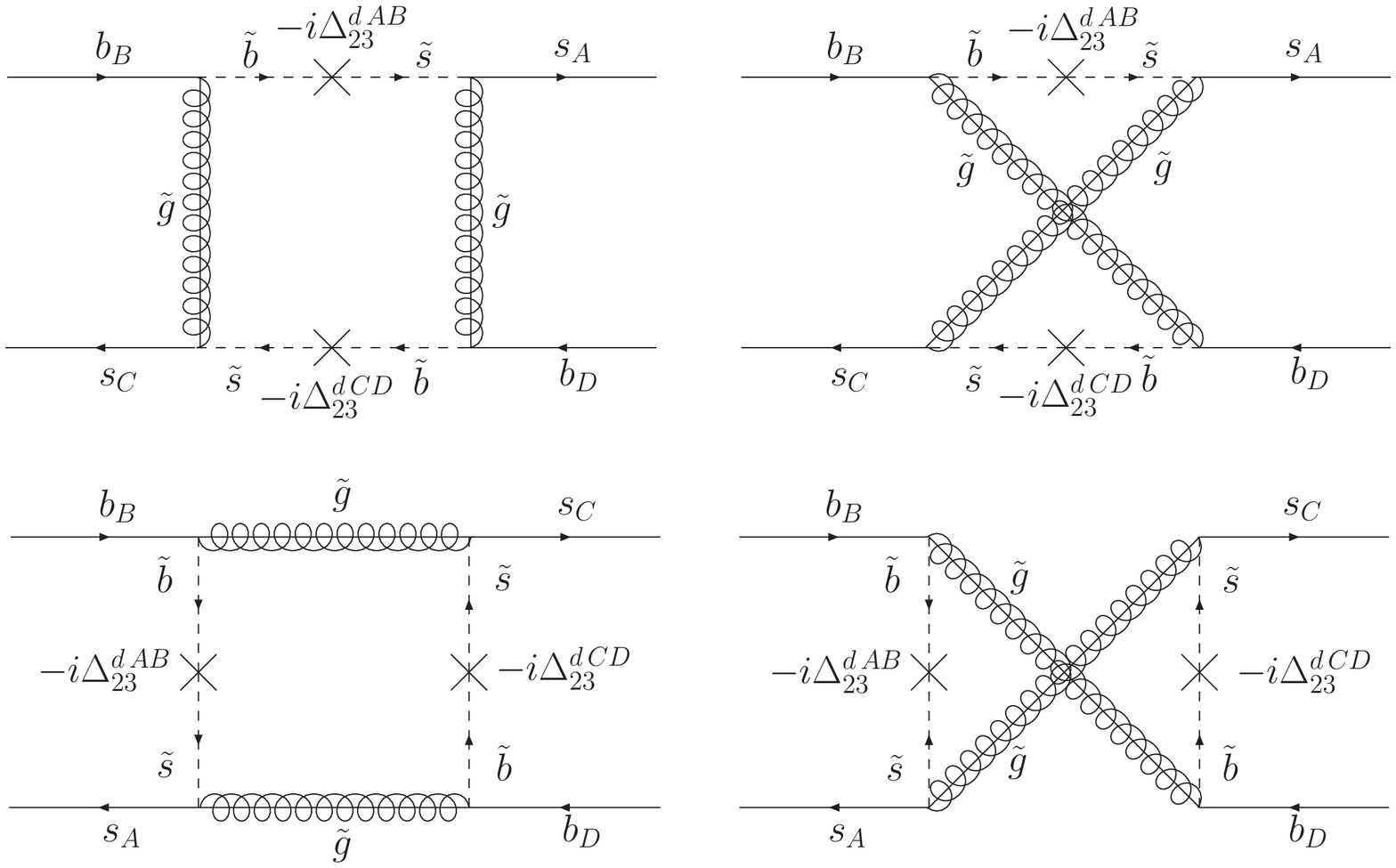}
\caption{One-loop gluino squark diagrams contributing to 
 \bbms\ in the presence of flavor-off-diagonal $A^d$ terms. 
 The crosses denote mass insertions while $A,B,C$ and $D$ label the 
 chiralities (equal to $L$ or $R$) with $A\neq B$ and $C\neq D$.
 \label{Gluino-Boxen}}
\end{nfigure}

Large FCNC effects with a different phenomenological pattern can be expected
if the squark mass matrices contain large chirality-flipping but
flavor-conserving entries. This situation occurs if the ratio
$\tan\beta=v_u/v_d$ of the Higgs vevs is large and can be traced back to an
effective loop-induced coupling of the Higgs doublet $H_u$ to down-type quarks
$d_j$, where $j=1,2,3$ labels the fermion generation \cite{Banks:1987iu}.
Hall, Rattazzi and Sarid discovered the relevance of this loop contribution
for large-$\tan\beta$ phenomenology \cite{Hall:1993gn} and the authors of
Ref.~\cite{Hamzaoui:1998nu} observed that $\tan\beta$-enhanced loop-induced
FCNC Higgs couplings occur even in MFV scenarios.  If the chirality violation
is proportional to a Yukawa coupling further a resummation of the enhanced
loops to all orders in perturbation theory is necessary. This resummation can
be achieved analytically with the help of an effective two-Higgs-doublet
Lagrangian (valid in the decoupling limit of infinite sparticle masses)
\cite{Hamzaoui:1998nu, twohdm} or through an explicit diagrammatic resummation
\cite{Carena:1999,Marchetti:2008hw,Hofer:2009xb}. In the diagrammatic method
the enhanced loop effects occur through chirality-flipping self-energies which
are chirally enhanced by a factor of $\tan\beta$.

In this paper we study the effects of chirally-enhanced flavor-changing
self-energies in the generic MSSM.     
The first possibility for a chiral enhancement factor
is $A^d_{ij}v_d/(M_{\rm SUSY}\, {\rm Max}[m_{d_i},m_{d_j}])$ with a
flavor-changing trilinear SUSY-breaking term $A^d_{ij}$ dominating over
a small quark mass $m_{d_{i,j}}$ in the denominator. $M_{\rm SUSY}$ is
the mass scale determining the size of the loop diagram, i.e. $M_{\rm{SUSY}}$ is roughly the maximum of the gluino and squark masses running
in the loop.  The second possible chiral enhancement factor is
$(v/M_{\rm SUSY})\, \tan\beta$ accompanied by a flavor-changing squark
mass term involving squark fields of the same chirality.  These
self-energies have been analyzed in the context of charged-current
processes in Ref.~\cite{Crivellin:2008mq} and this paper contains the
complementary study of FCNC processes. As an example, consider \bbms,
with the LO diagrams shown in \fig{Gluino-Boxen}. In the presence of
chirally enhanced corrections one must also take into account two- or
even tree-loop diagrams, because the loop suppression is offset by the
chiral enhancement factor (see \fig{Gluino-Boxen+Selbstenergien}).
\begin{nfigure}{t}
\includegraphics[width=0.9\textwidth]{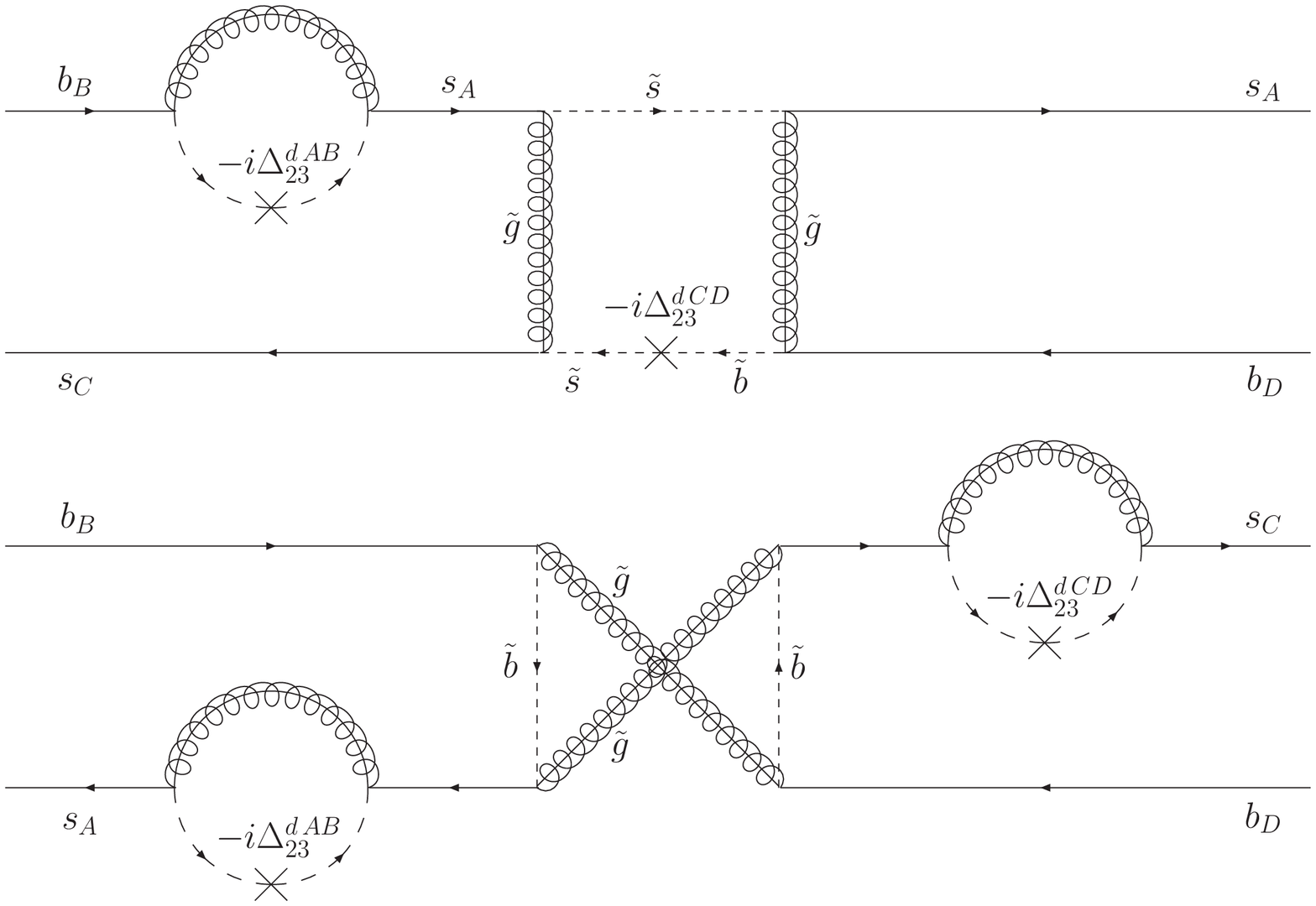}
\caption{Examples of two-loop and three-loop diagrams which can compete with (or even dominate over)
  the one-loop diagrams in \fig{Gluino-Boxen}.
  \label{Gluino-Boxen+Selbstenergien}}
\end{nfigure}
Similar corrections have been considered before in
Ref.~\cite{Foster}.  However, the authors of these papers have used a
different definitions of the super-CKM basis and of the parameters $\delta_{fi}^{q\;AB}$
describing the flavour violation in the squark mass matrices. 
As a consequence, our results are hardly comparable to the ones obtained in Ref.~\cite{Foster}. We elaborate on the
differences between Ref.~\cite{Foster} and this paper in
Sect.~\ref{sect:2}. We note that models in which CKM elements and
light-fermion masses are generated radiatively
\cite{Borzumati:1997bd,Crivellin:2008mq} require large trilinear
SUSY-breaking terms. In the presence of these large $A$-terms (or of a
large factor of $m_b \mu \tan\beta$ in combination with
chirality-conserving flavor violation) it is important to include the
effect of the chirality-flipping self-energies into FCNC processes. We
will accomplish this in Sect.~2 by renormalizing the quark-squark-gluino
vertex by a matrix-valued quark-field rotation in flavor space.  In
Sect.~3 the radiative decay $b\to s\gamma$ is examined in detail. The
chirally enhanced corrections are only relevant for the
large-$\tan\beta$ case here (or if $v_d A^d_{33}$ large).  In Sect.~4,
$\Delta F=2$ processes are investigated, where large corrections occur
irrespective of the size of $\tan\beta$ if the flavor violation is due
to $A$-terms and if the squarks are not degenerate. Up-to-date
measurements and theoretical Standard-Model predictions are used. The
theoretical uncertainties of the SM are treated in a consistent and
systematic way.  In each case we compare the size of the FCNC transition
to the previously known LO result: With the inclusion of our
self-energies the bounds on the off-diagonal elements of the squark mass
matrix change drastically, especially if the SUSY particles are rather
heavy. In Sect.~5 we conclude.

\section{One-loop renormalization of the quark-squark-gluino vertex}\label{sect:2}%
 
In this section we show how to treat the chirally enhanced parts of
flavor-changing self-energies in the full MSSM and how to absorb them
into mixing matrices entering the Feynman rule of the
quark-squark-gluino vertex. In this way all enhanced two-loop and
three-loop diagrams are automatically included in the LO calculation
to all orders in $v/M_{\rm SUSY}$.

Chirally enhanced corrections, in which we are interested here, have
been calculated in Ref.~\cite{Foster} from a loop-corrected quark mass
matrix, in a formalism in which the heavy SUSY particles are
integrated out. When these expressions are combined with some one-loop
amplitude this method can, in principle, reproduces
the chirally enhanced two-loop (or three-loop) FCNC amplitude to
leading non-vanishing order in $v/M_{\rm SUSY}$, where
$v=\sqrt{v_u^2+v_d^2}=174\,$GeV is the electroweak vev. Since the
squark mixing angles scale as $v/M_{\rm SUSY}$, scenarios with large
left-right mixing among squarks are not properly covered in this
approach.  In particular, the widely-studied large-$\tan\beta$
scenarios typically involve large sbottom mixing and therefore call
for an analysis beyond the decoupling limit. On the other hand,
Ref.~\cite{Foster} goes beyond our work by also including electroweak
and Higgs-mediated contributions to the Wilson-coefficients.  Another
important difference between Ref.~\cite{Foster} and this paper
concerns the definitions of the super-CKM basis and the mass insertion
parameters:
\begin{itemize}
	\item We use the tree-level definition of the super-CKM basis (see
Ref.~\cite{Crivellin:2008mq}) which permits an analytical solution 
to the necessary all-order resummations of enhanced corrections. 
With the on-shell definition of Ref.~\cite{Foster} the self-energies
and the squark mass matrices depend mutually on each other, which
requires an iterative numerical evaluation of both quantities. This definition clouds the relations between
observables and fundamental parameters like the trilinear
SUSY-breaking terms $A_{fi}^{u,d}$ (as discussed later in this section).
Our results are more transparent and enable us to identify a previously
neglected parameter region with dramatically enhanced corrections to
meson-antimeson mixing (see Sect.~\ref{sect:dfp}.)
\item Concerning the definition of the mass insertion parameters
$\delta_{fi}^{q\;AB}$ we choose the most common one (see for example
\cite{Gabbiani:1996,Ciuchini:1997bw,Ciuchini:1998ix,Borzumati:1999,Becirevic:2001,Ciuchini:2002uv,Ciuchini:2006dx,Ciuchini:2006dw,Ciuchini:2007cw,Crivellin:2008mq})
in which the mass insertion is the entire off-diagonal element of the
squark mass matrix divided by the average squark mass. On the other
hand, the authors of Ref.~\cite{Foster} only include
the term $v_q A^q$ into their definition of the mass insertion
parameters. This leads to an artificial dependence of all corrections,
even the ones independent of a quark mass, on
$\tan\beta$. Furthermore, if the authors of Ref.~\cite{Foster} would
have chosen our definition of the mass insertions (while keeping their
definition of the super-CKM basis) all chirally enhanced corrections
would be simply absent, instead they would be
implicitly contained in the definition of the
$\delta_{fi}^{q\;AB}$'s.
   
\end{itemize}

Throughout this paper we use the notations and conventions of
Ref.~\cite{Crivellin:2008mq}. The self-energies can be divided into a
chirality-flipping and a chirality-conserving part:
\begin{equation}
\Sigma _{fi}^q (p) = 
\Sigma^{q\,RL}_{fi} \left( {p^2 } \right) P_L  + 
\Sigma^{q\,LR}_{fi} \left( {p^2 } \right) P_R  + 
\ps  \left[ \Sigma^{q\,LL}_{fi} \left( {p^2 } \right) P_L  + 
            \Sigma^{q\,RR}_{fi} \left( {p^2 } \right) P_R \right]
\label{Zerlegung}
\end{equation}
Since the SUSY particles are known to be much heavier than the five lightest
quarks, it is possible to expand in the external momentum unless one external
quark is the top. We are not going to study the case of an external top
further in this article, because there are no useful data on FCNC top decays.
The part proportional to $\ps$ in \eq{Zerlegung} can be
neglected since no chiral enhancement is possible. Thus the following
simplified expression for the self-energy is sufficient for our purpose:
\begin{equation}
\Sigma^{q\,RL,LR}_{fi} (p^2=0) =
\frac{{2m_{\tilde g} }}{{3\pi }}\alpha _s (M_{\rm SUSY})
      \sum_{s = 1}^6 V^{(0)\,q\, RL,LR}_{s\,fi} 
  B_0 \left( {m_{\tilde g} ,m_{\tilde q_s } } \right),
\label{selbstenergie}
\end{equation}
If $v_q A^q_{fi}$ is large, FCNC diagrams with this self-energy in an external
leg (as in \fig{Gluino-Boxen+Selbstenergien}) will compete with the leading
order diagram (see \fig{Gluino-Boxen}). For $q=d$ the enhancement can further
stem from $m_b \mu\tan\beta$ in which case an all-order resummation 
is necessary \cite{Carena:1999,Hofer:2009xb}, leading to a correction 
of the tree-level relation $m_{d_i}=v_d Y^{d_i}$ to 
\begin{equation}
Y^{d_i} \quad =\quad \frac{m_{d_i}}{v_d \, (1 + \Delta_{d_i}) } 
       \quad = \quad  
         \frac{m_{d_i}-\Sigma^{d\,LR}_{ii,\,A}}{
               v_d \, + \Sigma^{d\,LR}_{ii,\,\mu}/Y^{d_i}  } 
        .\label{mqdel}
\end{equation}
In \eq{mqdel} we have used the fact that $\Sigma^{d\,LR}_{ii}$ can be
decomposed into $\Sigma^{d\,LR}_{ii,\,A}+\Sigma^{d\,LR}_{ii,\,\mu}$ even if
the physical squark masses are chosen as input parameters.
$\Sigma^{d\,LR}_{ii,\,\mu}$ is proportional to $\mu\, Y_{d_i}$ (and
$\Sigma^{d\,LR}_{ii,\,\mu}/Y^{d_i}$ is independent of $Y^{d_i}$) and
$\Sigma^{d\,LR}_{ii,\,A}$ is proportional to $A^{d}_{ii}$. If we neglect
the latter contribution \eq{mqdel} reduces to the expression of
Ref.~\cite{Carena:1999}. $m_{d_i}$ is the physical quark mass, with an on-shell
subtraction of the SUSY contributions, as determined from low-energy data.
For a detailed discussion of the relation between Yukawa couplings and quark
masses for different renormalization schemes see Ref.~\cite{Hofer:2009xb}.
For up-quarks one must replace $d\to u$ in \eq{mqdel} and may omit
$\Sigma^{d\,LR}_{ii,\,\mu}$, which is now suppressed with 
$\cot\beta$.

As already mentioned, the chirally enhanced parts of the SQCD self-energies do
not decouple. This means they do not vanish in the limit $M_{\rm SUSY}\to
\infty$ but rather converge to a constant. Therefore the corresponding FCNC
diagrams of the type in \fig{Gluino-Boxen} scale with $M_{\rm SUSY}$ in the
same way as the LO diagram. We will find that the two-loop and three-loop
diagrams compete with the LO ones, especially for rather large and
non-degenerate squark masses. There are different possibilities to handle
flavor-changing self-energies in external legs: The conceptually simplest
version is to treat them in the same way as one-particle-irreducible diagrams
\cite{ln}. So far we have implied this method, which is best suited to
identify chirally enhanced contributions.  Calculations are easiest, however,
if one absorbs the chirally enhanced corrections into the quark-squark-gluino
vertex through unitary rotations $\Delta U^{q}_{A}$ ($A=L,R$ denotes the
chirality) in flavor space \cite{Crivellin:2008mq}.  These rotations alter 
the Feynman rule for the squark-quark-gluino vertex:
\begin{equation}
\begin{array}{l}
 W_{s,i}^{q*}  \to W_{s,j}^{q*} \left( {1 + \Delta U_{L\;ji}^q } \right) \\ 
 W_{s,i + 3}^{q*}  \to W_{s,j + 3}^{q*} 
 \left( {1 + \Delta U_{R\;ji}^q } \right) \\ 
 \end{array}
 \label{Wren}
\end{equation}
with
\begin{equation}
\renewcommand{\arraystretch}{1.4}
\begin{array}{l}
\Delta U_L^q \,=\, 
\left( {\begin{array}{*{6}c}
0 & 
{\frac{1}{{m_{q_2 }}} {\Sigma _{12}^{q\,LR} }  } & 
{\frac{1}{{m_{q_3 }}} {\Sigma _{13}^{q\,LR} } }  \\
           {\frac{{ - 1}}{{m_{q_2 }}} {\Sigma _{21}^{q\,RL} }  } & 
0 & 
{\frac{1}{{m_{q_3 }}} {\Sigma _{23}^{q\,LR} } }  \\
           {\frac{{ - 1}}{{m_{q_3 }}} {\Sigma _{31}^{q\,RL} } } & 
{\frac{{ - 1}}{{m_{q_3 }}}{\Sigma _{32}^{q\,RL} }  } & 0
        \end{array}} \right) \\ \\[-3mm]
\Delta U_R^q \,=\, 
\left( {\begin{array}{*{6}c}
0 & 
{\frac{1}{{m_{q_2 }}} {\Sigma _{12}^{q\,RL} }  } & 
{\frac{1}{{m_{q_3 }}} {\Sigma _{13}^{q\,RL} } }  \\
           {\frac{{ - 1}}{{m_{q_2 }}} {\Sigma _{21}^{q\,LR} }  } & 
0 & 
{\frac{1}{{m_{q_3 }}} {\Sigma _{23}^{q\,RL} } }  \\
           {\frac{{ - 1}}{{m_{q_3 }}} {\Sigma _{31}^{q\,LR} } } & 
{\frac{{ - 1}}{{m_{q_3 }}}{\Sigma _{32}^{q\,LR} }  } & 0
        \end{array}} \right)\\        
 \end{array}       
\label{DeltaU} 
\end{equation}
The procedure in \eq{Wren} can be viewed as a short-cut to include the
self-energy in the external quark line, in the spirit of Ref.~\cite{ln}.
Alternatively \eq{Wren} can be interpreted as a finite matrix-valued
renormalization of the quark fields which cancels the external self-energies
and reappears in the Feynman rule of the quark-squark-gluino
vertex.\footnote{We stress that we do not introduce ad-hoc counter-terms to the
  quark-squark-gluino vertex. Supersymmetry links the renormalization of the
  latter to the quark-quark-gluon vertex (which is unaffected by the rotations
  in \eq{DeltaU}) and the renormalization of the soft SUSY-breaking terms
  (which can feed into the renormalization of the squark rotation matrices).
  The requirement to maintain the structure of a softly broken SUSY theory
  within the renormalization process restricts the allowed counterterms to the
  quark-squark-gluino vertex. Counterterms stemming from field
  renormalizations, however, are harmless in this respect, because field
  renormalizations trivially drop out from the LSZ formula for transition
  matrix elements.}%
The inclusion of the enhanced corrections into the LO calculation is now
simply achieved by performing the replacements of \eq{Wren} in this Feynman
rule.  Therefore here the exact diagonalization of the squark mass matrix is
preferred over MIA. The exact diagonalization has also the advantage that the
analysis can be extended to the large $\tan\beta$ region in which certain
off-diagonal entries can have the same size as the diagonal ones.

Here a comment on the definition of the super-CKM basis and the
renormalization scheme is in order (see also Ref.~\cite{Crivellin:2008mq}):
\begin{itemize}
\item[i)] We define the super-CKM basis as follows: Starting from some weak
  basis we diagonalize the tree-level Yukawa couplings and apply this unitary
  transformation to the whole supermultiplet. This is a natural definition of
  the super-CKM basis because of the direct correspondence between the
  SUSY-breaking Lagrangian and physical observables. Whenever we refer to 
  some element of a squark mass matrix, this element is defined in this 
  basis. When passing from LO to NLO or even higher orders the definition 
  of the squark mass matrices is unchanged, i.e.\ no large 
  chirally enhanced rotations appear at this step. Any additional 
  non-enhanced (i.e.\ ordinary SQCD) corrections are understood to be
  renormalized in a way which amounts to a minimal  renormalization of 
  the squark mass matrices.
\item[ii)] From given squark mass matrices we calculate the self-energies
  $\Sigma _{fi}^{q\,RL}$ and then the rotations $\Delta U_{L,R}^q$ in
  \eq{Wren}.  Calculating LO amplitudes with the corrected $W_{sk}^{q*}$ from
  \eq{Wren} then properly includes the desired chirally enhanced effects. 
  The order of the two steps is important: First the  super-CKM basis is 
  defined from the tree-level structure of the Yukawa sector and the finite 
  loop effects are included afterward, without influence on the definition 
  of the super-CKM basis.  
\end{itemize}
Alternatively one could define the super-CKM basis using an on-shell scheme
which eliminates the self-energies in the external legs by shifting their
effect into the definition of the super-CKM basis: Applying the inverse of the
rotation in \eq{Wren} first to the whole (s)quark superfields will leave the
squark-quark-gluino vertex flavor-diagonal. (Further supersymmetry is still
manifest, e.g.\ the sbottom field is the superpartner of the bottom field.
This would not be the case if different rotations were applied to quark and
squark fields.)  If one defines this basis as the super-CKM basis (which now
changes in every order of perturbation theory) one will find very different
constraints on the off-diagonal elements of the squark mass matrices than with
our method. The effect of the enhanced self-energies will be entirely absorbed
into the values of the elements of the squark mass matrices, these
self-energies will not appear explicitly, and the calculation of LO diagrams
in the usual way will be sufficient. However, the squark mass matrices
determined from data using this method will not be simply related to a
mechanism of SUSY breaking, because the extracted numerical value of a given
matrix element will also contain the physics associated with the chirally
enhanced self-energies. Effects from SUSY breaking and electroweak breaking 
are interwoven now and further the elements of the squark mass matrix do not
obey simple RG equations anymore.  

It is illustrative to consider the popular case of soft-breaking terms which
are universal at a high scale, say, the GUT scale.  The unitary rotations
diagonalizing the Yukawa couplings will lead to soft-breaking terms which are
proportional to the unit matrix in flavor space.  The RG evolution down to low
scales will then lead to small flavor-off-diagonal LL entries of the squark
mass matrices which are governed by the tree-level (as defined in
Ref.~\cite{Crivellin:2008mq}) CKM matrix.  Obviously, the elements of the
squark mass matrices defined in this way are the quantities which one wants to
probe in order to discriminate between high-scale universality and other
possible mechanisms of soft flavor violation.  The above-mentioned unitary
rotations diagonalize the tree-level Yukawa couplings, while the rotations
with $\Delta U_{L,R}^q$ are only meaningful after electroweak symmetry
breaking and are therefore a low-scale phenomenon.  During the RG evolution
the Yukawa couplings essentially stay diagonal and the small unitary rotations
bringing the low-scale Yukawa couplings back to diagonal form are unrelated to
the soft breaking sector (and involve no chiral enhancement).  Therefore the
procedure described in items i) and ii) is the adequate method to probe the
flavor structure of SUSY breaking.  For a discussion of renormalization
schemes in the context of MFV see Ref.~\cite{Degrassi:2006eh}.

Consider an FCNC transition $d_i\to d_f$ in MIA: If the squark mass $m_{\tilde
  d_i }$ is degenerate with $m_{\tilde d_f }$, the renormalization effects of
the squark-quark-gluino vertex drop out in FCNC processes. This can be
understood in the diagrammatical approach by realizing that diagrams with a
flavor-changing self-energy in the outgoing $f$ leg cancel with the diagram
where the self-energy is in the incoming $i$ leg, because the intermediate
loop is the same for both diagrams.  Thus, in order to demonstrate the effects
of the renormalized quark-squark-gluino vertex non-degenerate squarks are
necessary.  For definiteness we choose the flavor-diagonal left-handed and
right-handed mass terms equal and further set $m_{\tilde{q}1,\tilde{q}2}=
2m_{\tilde{q}_3}$ if not mentioned otherwise. Lighter third-generation squarks
are plausible in scenarios with high-scale flavor universality, in which
renormalization group (RG) effects usually drive the bilinear soft terms of
the third generation down.

\begin{nfigure}{t}
  \includegraphics[width=0.48\textwidth]{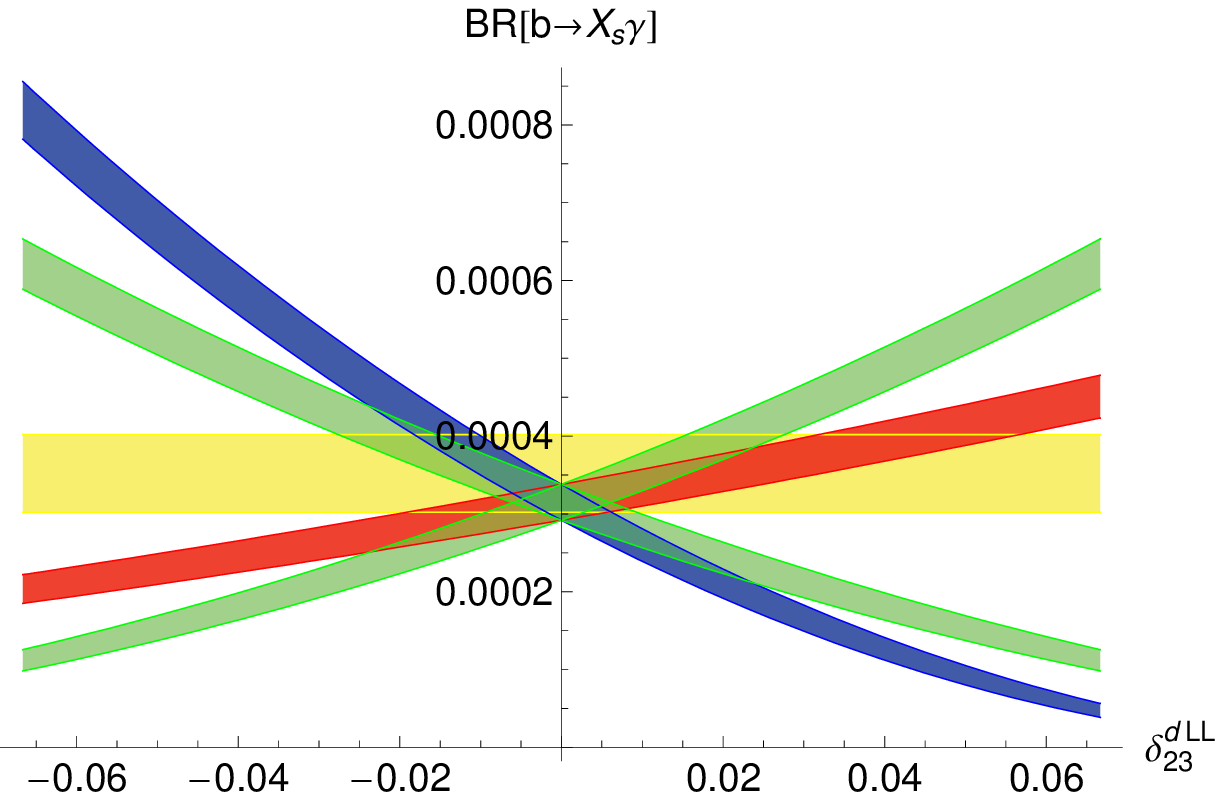} ~~~
  \includegraphics[width=0.48\textwidth]{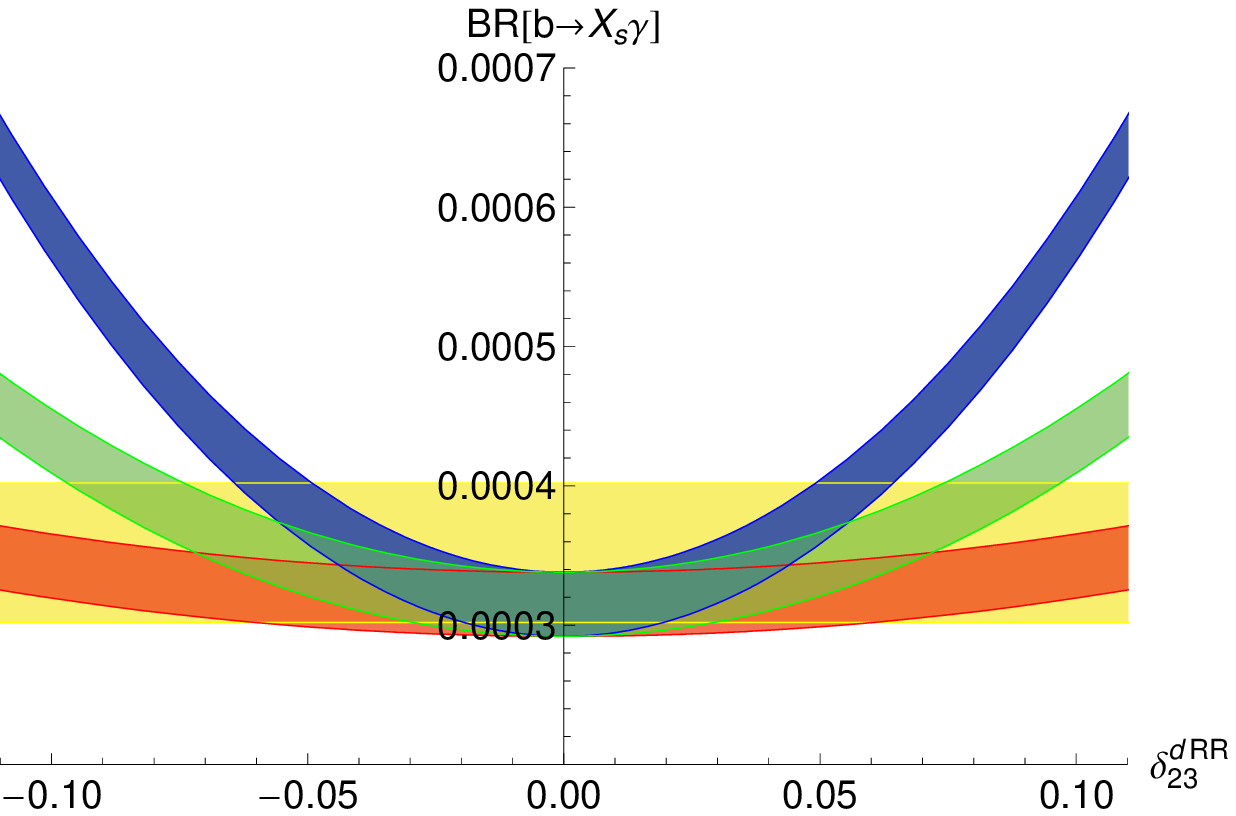}
\caption{$\rm{Br}[B\to\rm{X_s}\gamma]$ 
  as a function of $\delta^{d\,LL}_{23}$ and $\delta^{d\,RR}_{23}$,
  respectively, for $m_{\tilde{g}}=750\,\textrm{GeV}$,
  $m_{\tilde{q}1/2}=2m_{\tilde{q}3}=1000\,\textrm{GeV}$ and $\tan\beta=50$.
  Yellow(lightest): experimentally allowed range for
  $\rm{Br}[B\to\rm{X_s}\gamma]$. Green, red, blue (light to dark):
  theoretically predicted range for $\mu/(1+\Delta_b)=\pm 600\,\textrm{GeV}$
  without renormalization, $+600\,\textrm{GeV}$ with renormalized vertices,
  $-600\,\textrm{GeV}$ with renormalized vertices. In the left figure the
  green band with positive (negative) slope corresponds to the LO
  branching-ratio with $\mu/(1+\Delta_b)=+ 600\,\textrm{GeV}$
  ($\mu/(1+\Delta_b)=- 600\,\textrm{GeV}$).
  \label{b-s-gammaLL/RR}}
\end{nfigure}

At this point we may compare our results in
\eq{DeltaU} with the corresponding expressions in Ref.~\cite{Foster}.
Unlike our $\Delta U_{L,R}^q$ the enhanced corrections in
Ref.~\cite{Foster} depend explicitly on $\tan\beta$. This feature reflects the
different definitions of the $\delta_{fi}^{q\;AB}$'s adopted in the
two approaches. In particular, the constraints which we will derive 
from $\Delta F=2$ processes in Sect.~\ref{sect:dfp} are very different
from those in Ref.~\cite{Foster}.

\boldmath
\section{$B\rightarrow X_s\gamma$}
\unboldmath%

In this section we examine the radiative decay
$b\rightarrow s\gamma$ . We show how the renormalization of the
quark-squark-gluino vertex affects the branching-ratio for different
values of $\mu$.\footnote{We find agreement 
{of our LO}
Wilson coefficients {with} the gluino part given in
Ref.~\cite{Borzumati:1999,Besmer:2001cj}.} 
Throughout this section we assume that $\mu$ and the elements
$\Delta^{q\,AB}_{ij}$ with $AB=LL,LR,RL,RR$, of the squark mass
matrices are real. We consider only the case in which $|\mu| \tan
\beta$ is large, because otherwise our new contributions are
suppressed and no new constraints can be found. The reason for this
feature is the chirality structure of the magnetic transition mediated
by the operator $O_7$: Both the flavor-changing self-energy and the
now flavor-conserving magnetic loop are necessarily
chirality-flipping. The chiral enhancement of the latter is achieved
by a large value of $|\mu| \tan \beta$. We further recall that the
contributions from the dimension-six operators $O_{7b,\tilde{g}}$ and
$O_{8b,\tilde{g}}$ (defined according to Ref.~\cite{Borzumati:1999})
are suppressed by a factor of $\frac{M_{\rm SUSY}}{\mu \tan\beta}$
compared to the contributions from $O_{7\tilde{g}}$ and
$O_{8\tilde{g}}$.  Furthermore, since all other SQCD contributions are
also suppressed we only need to consider the magnetic operators and
their chromomagnetic counterparts:
\begin{equation}
O_{7\tilde{g}} = e g_s^2(Q)\, \bar{s}\sigma_{\rho\nu} P_R 
                 b\, F^{\rho\nu},\qquad\qquad
O_{8\tilde{g}}= g_s^3 (Q) \bar{s}\sigma_{\rho \nu} 
                  T^a P_R b\, G^{a\rho\nu}
\end{equation}
\begin{equation}
 \tilde{O}_{7\tilde{g}} = e g_s^2(Q)\, 
   \bar{s}\sigma_{\rho\nu}P_L b\, F^{\rho\nu}, \qquad\qquad
\tilde{O}_{8\tilde{g}}=g_s^3(Q)\, 
   \bar{s} 
    \sigma_{\rho\nu}T^a P_L b\, G^{a\rho\nu}               
\end{equation}
In \cite{Borzumati:1999} the matching scale $Q$ is chosen as $Q=M_W$. We use
$Q=m_t$ instead, because it is closer to the SUSY scale while still permitting
5-flavor running of $\alpha_s$.  The experimental value of
\cite{Amsler:2008zzb} is taken at $2\sigma$ confidence level. For the
theoretical prediction, the value of reference \cite{Misiak:2006ab} is used at
the lower and upper end of the error range. We have not used the cumbersome 
NNLO formula of Ref.~\cite{Misiak:2006ab}, but have instead fitted  $C_{7SM}$ 
in the simple LO formula to reproduce the numerical NNLO result
for $\Gamma \left( {b \to s\gamma } \right)$. The LO expression reads
\begin{equation}
\renewcommand{\arraystretch}{2.4}
\begin{array}{c}
 \Gamma \left( {b \to s\gamma } \right) = 
\displaystyle{\frac{{m_b^5 G_F^2 \left| {V_{tb}^{} V_{ts}^* } \right|^2 
                    \alpha }}{
             {32\pi ^4 }}\left( {\left| {C_7} \right|^2  + 
                            \left| {\tilde C_7} \right|^2 } \right) }\\ 
 C_7 = \displaystyle{\frac{{ - 16\sqrt 2 \pi ^3 \alpha _s 
           \left( {\mu _b } \right)}}{{G_F V_{tb} V_{ts}^* m_b }}
 C_{7\tilde g}  + C_{7SM}}  \\ 
 \tilde C_7 = \displaystyle{\frac{{ - 16\sqrt 2 \pi ^3 \alpha _s 
    \left( {\mu _b } \right)}}{{G_F V_{tb} V_{ts}^* m_b }}
     \tilde C_{7\tilde g}}  \\ 
 \end{array}
\label{Verzweigunsverhaeltnis}
\end{equation}
To check our approximations we have 
also calculated the NLO evolution with matching at $Q=M_{\rm SUSY}$, but 
found only a slightly different result.

We now discuss the dependence of $b\to s\gamma$ on the different squark mass
parameters $\Delta^{d\,AB}_{23}$ (or, equivalently, on the usual dimensionless
quantities $\delta^{d\,AB}_{23}$): If the chirality-conserving elements
$\Delta^{d\,LL,RR}_{23}$ are the non-minimal source of flavor violation $b\to
s\gamma$ depends very strongly on $\mu\tan\beta$ already at the one-loop level
(i.e.\ without the renormalization of the quark-squark-gluino vertex). With
the inclusion of the flavor-changing self-energies in the external legs
$C_{7\tilde{g},8\tilde{g}}$ is enhanced (suppressed) if $\mu$ is negative
(positive) compared to the LO coefficient.  The size of the effect is rather
different for $\delta^{d\,LL}_{23}$ and $\delta^{d\,RR}_{23}$, because only in
the first case interference with $C_{7SM}$ is possible (see
\fig{b-s-gammaLL/RR}).

For the chirality-violating elements of the squark mass matrix,
$\Delta^{d\,LR,RL}_{23}$, this dependence on $\mu\tan\beta$ is absent at LO
and comes only into the game by the renormalization of the squark-quark-gluino
vertex. Again the behavior is different for $\delta^{d\,LR}_{23}$ compared to
$\delta^{d\,RL}_{23}$, since only in the first case interference with
$C_{7SM}$ is possible (see \fig{b-s-gammaLR/RL}).
\begin{nfigure}{tb}
\includegraphics[width=0.48\textwidth]{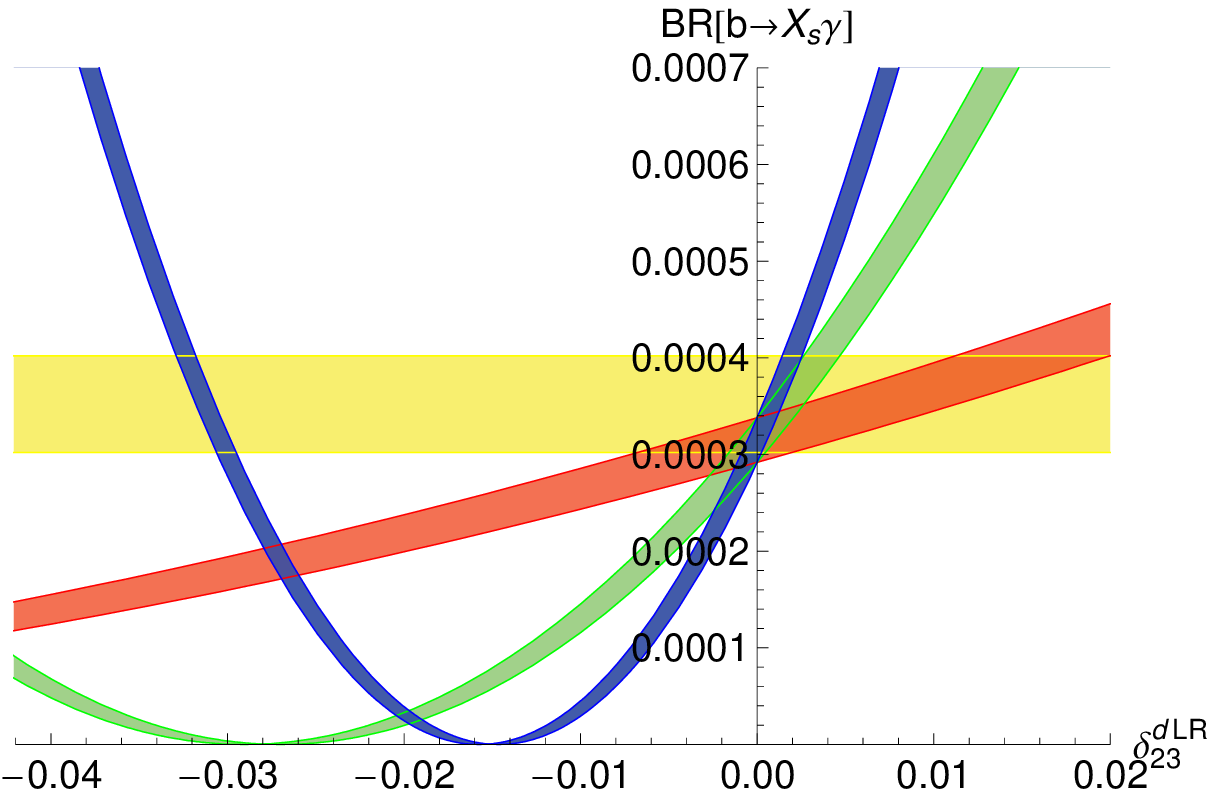}
~~~
\includegraphics[width=0.48\textwidth]{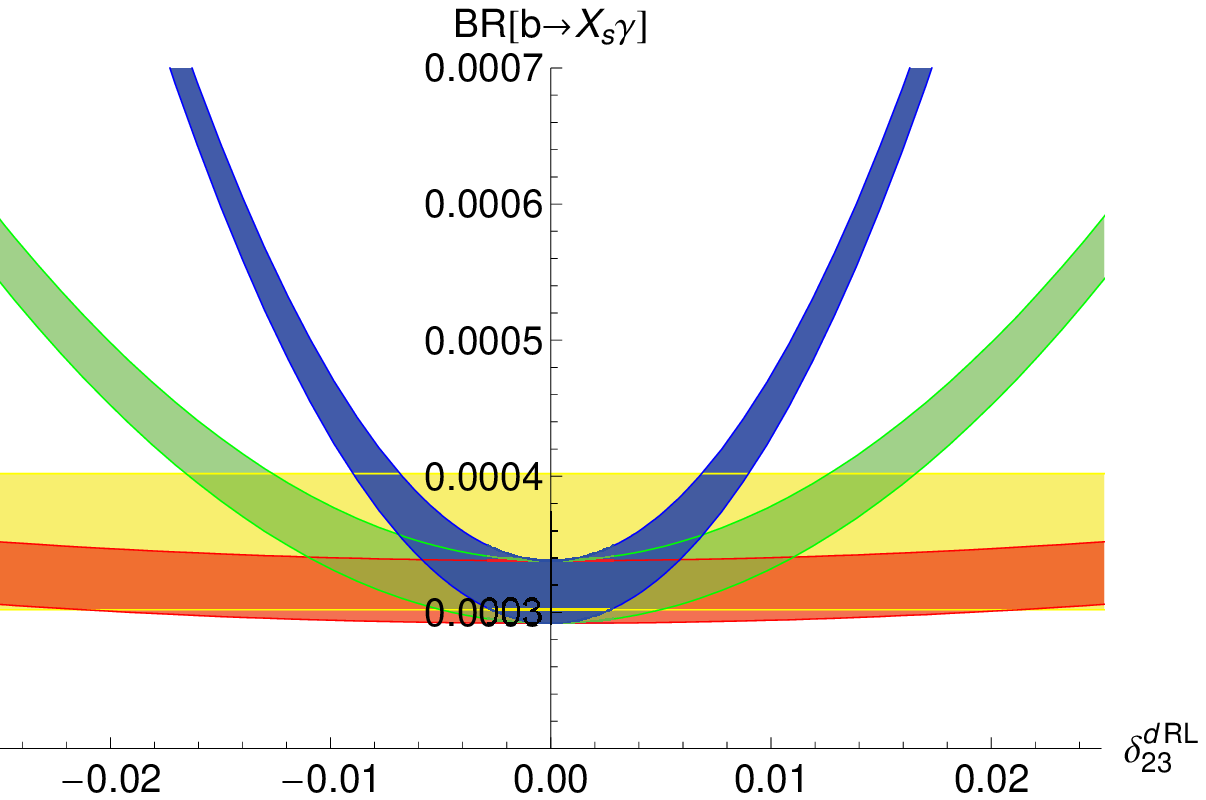}
\caption{$\rm{Br}[B\to\rm{X_s}\gamma]$  
  as a function of $\delta^{d\,LR}_{23}$ and $\delta^{d\,RL}_{23}$,
  respectively, for $m_{\tilde{g}}=750\,\textrm{GeV}$,
  $m_{\tilde{q}1,\tilde{q}2}=2m_{\tilde{q}3}=1000\,\textrm{GeV}$, and 
  $\tan\beta=50$.
  Yellow(lightest): experimentally allowed range for
  $\rm{Br}(B\to\rm{X_s}\gamma)$. Green, red, blue (light to dark):
  theoretically predicted range for $\mu/(1+\Delta_b)=0\,\textrm{GeV}$,
  $\mu/(1+\Delta_b)=600\,\textrm{GeV}$, $\mu/(1+\Delta_b)=-600\,\textrm{GeV}$.
  \label{b-s-gammaLR/RL}}
\end{nfigure}
As we easily see from \fig{b-s-gammaLR/RL} the inclusion of the two-loop
effects can substantially change the branching ratio.  For positive (negative)
values of $\mu$ the size of the Wilson coefficient $C_{7\tilde{g}}$ decreases
(increases), i.e.\ the qualitative effect of our corrections is the same 
as for the LL and RR elements in \fig{b-s-gammaLL/RR}. 

In Fig.~\ref{b-s-gamma-constraints} we plot the constraints obtained from
$\rm{Br}[B\to\rm{X_s}\gamma] $ on $\delta^{d\,AB}_{23}$ versus
$\mu/(1+\Delta_b)$ for different values of $m_{\tilde{g}}$.
($\Delta_b=\Delta_{d_3}$ is defined in \eq{mqdel}.)  All four different
chirality combinations are shown. The constraints on $\delta^{d\,LR}_{23}$ and
$\delta^{d\,LL}_{23}$ are stronger than the ones on $\delta^{d\,RL}_{23}$ and
$\delta^{d\,RR}_{23}$ with exception of the "conspiracy" regions where the
SUSY contributions overcompensate the SM value for $C_7$. For
$\delta^{d\,LR,RL}_{23}$ the allowed region widens from bottom to top, meaning
that negative values of $\mu$ strengthen the bounds on these quantities while
positive values of $\mu$ weaken them. For $\delta^{d\,LL,RR}_{23}$ the effect
of $\mu$ is different, as one can verify from the two lower plots in
\fig{b-s-gamma-constraints}: The bounds on $\delta^{d\,LL,RR}_{23}$ always get
stronger for increasing $|\mu|$ but are more stringent for $\mu<0$ than for
$\mu>0$.
\begin{nfigure}{!tb}
  \includegraphics[width=0.48\textwidth]{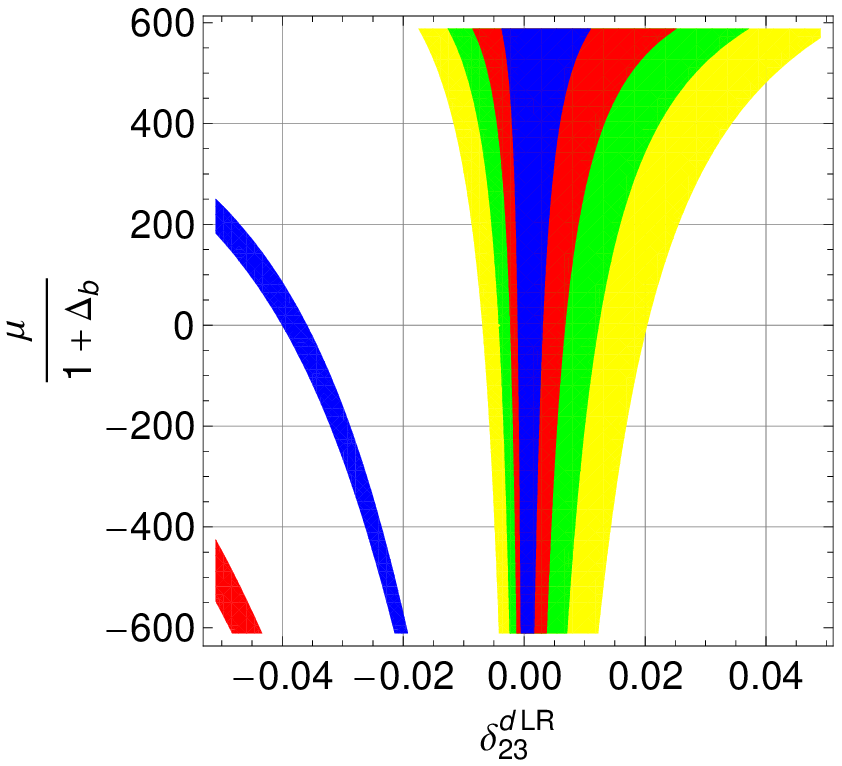} ~~~
  \includegraphics[width=0.48\textwidth]{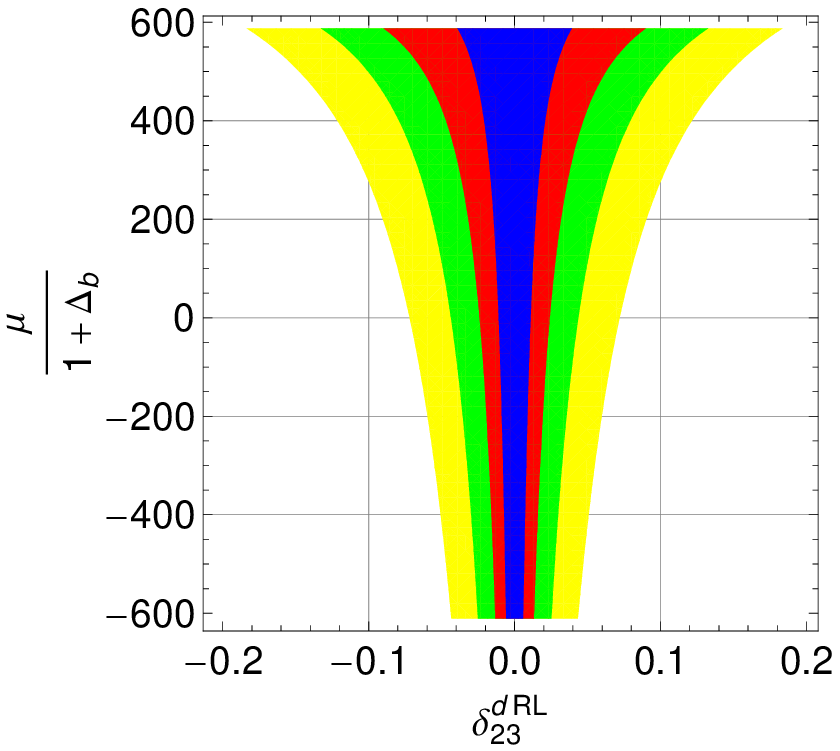}
  \includegraphics[width=0.48\textwidth]{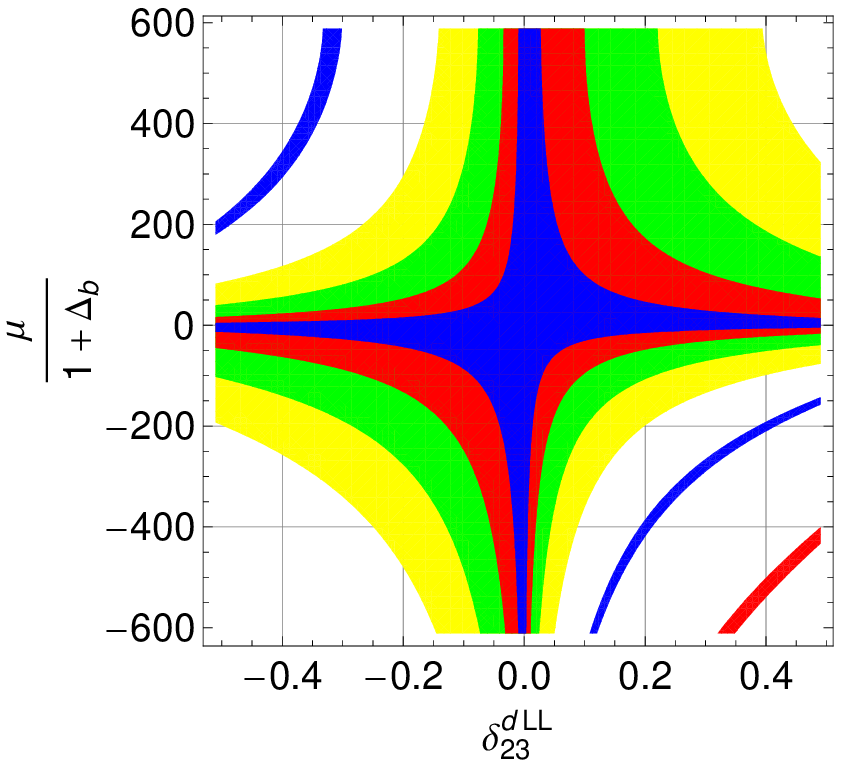} ~~~
  \includegraphics[width=0.48\textwidth]{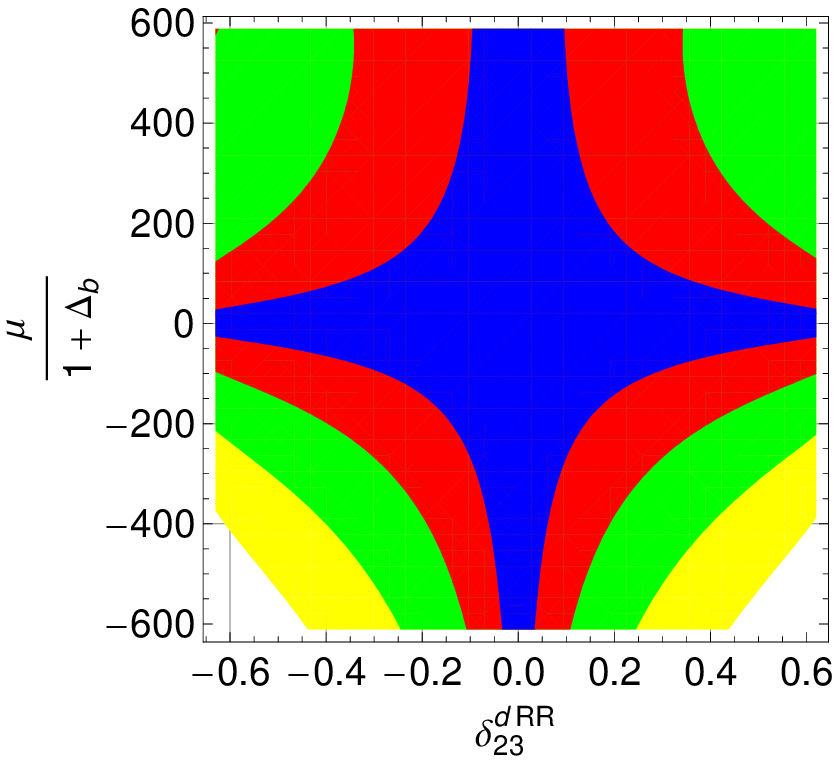}
\caption{$B\to\rm{X_s}\gamma$: allowed regions in the 
  $\delta^{d\,LR,RL}_{23}-\frac{\mu}{1+\Delta_b}$ and 
  $\delta^{d\,LL,RR}_{23}-\frac{\mu}{1+\Delta_b}$ planes for
  $m_{\tilde {q}1,2}=2m_{\tilde q3}=1000\,\textrm{GeV}$ and 
  $\tan\beta=50$. Yellow:
  $m_{\tilde{g}}=2000\,\textrm{GeV}$, green:
  $m_{\tilde{g}}=1500\,\textrm{GeV}$, red: $m_{\tilde{g}}=1000\,\textrm{GeV}$,
  blue: $m_{\tilde{g}}=500\,\textrm{GeV}$ (light to dark).
  \label{b-s-gamma-constraints}}
\end{nfigure}
\boldmath
\section{$\Delta F=2$ processes\label{sect:dfp}}
\unboldmath {In this section, we consider $B_d$, $B_s$,
and $K$ mixing. We show that the {enhanced} effects of
the renormalized quark-squark-gluino {vertices vanish
for degenerate} squark masses. However, if the squarks are
{non-degenerate our} (N)NLO corrections are even
dominant in a large region of the parameter space. In this analysis we
consider complex $\delta^{d\,AB}_{ij}$'s to exploit the data on CP
asymmetries.} The effective Hamiltonian is given by
\begin{equation}
\renewcommand{\arraystretch}{1.8}
\begin{array}{l}H_{\rm eff}^{\rm SUSY} = \\
 \begin{array}{@{}l}
  \displaystyle{\frac{- \alpha _s^2 }{216}} 
  \left[ \,\,\,\,\;V_{s\;23}^{d\;LL} V_{t\;23}^{d\;LL} 
  \left[ {24  m_{\tilde g}^2 D_0 
  \left( {m_{\tilde d_s } ,m_{\tilde d_t } ,m_{\tilde g} ,m_{\tilde g} } 
  \right) +  66 D_2 \left( {m_{\tilde d_s } ,m_{\tilde d_t } ,
        m_{\tilde g} ,m_{\tilde g} } \right)} \right]  Q_1\right.\\
   \;\;\;\;\;\;\;\;\,\;\;+ V_{s\;23}^{d\;RR} V_{t\;23}^{d\;RR} 
 \left[ {24 m_{\tilde g}^2  D_0  
 \left( {m_{\tilde d_s } ,m_{\tilde d_t } ,m_{\tilde g} ,m_{\tilde g} } 
    \right) + 
  66 D_2 \left( {m_{\tilde d_s } ,m_{\tilde d_t } ,
    m_{\tilde g} ,m_{\tilde g} } \right)} \right]\tilde Q_1 \\ 
  \;\;\;\;\;\;\;\;\,\;\;+ V_{s\;23}^{d\;LL} V_{t\;23}^{d\;RR} 
  \left[ {504 m_{\tilde g}^2 D_0 \left( {m_{\tilde d_s } ,m_{\tilde d_t } ,
      m_{\tilde g} ,m_{\tilde g} } \right) Q_4 - 
   72D_2 \left( {m_{\tilde d_s } ,m_{\tilde d_t } ,m_{\tilde g} ,
            m_{\tilde g} } \right)} Q_4 \right. \\ 
 \left. { \;\;\;\,\;\;\;\;\;\;\;\;\;\;\;\;\; 
          \;\;\;\;\;\;\;\;\;\;\;\;\;\;\;\;\;\, + 
    24m_{\tilde g}^2 D_0 \left( {m_{\tilde d_s } ,
          m_{\tilde d_t } ,m_{\tilde g} ,m_{\tilde g} } \right) Q_5 + 
   120D_2 \left( {m_{\tilde d_s } ,m_{\tilde d_t } ,
          m_{\tilde g} ,m_{\tilde g} } \right)} Q_5 \right] \\ 
  \;\;\;\;\;\;\;\,\;\;\;+V_{s\;23}^{d\;LR} V_{t\;23}^{d\;LR} 
  \left[ {204  m_{\tilde g}^2 D_0 \left( {m_{\tilde d_s } ,m_{\tilde d_t }
  ,m_{\tilde g} ,m_{\tilde g} } \right) Q_2 
    - 36 m_{\tilde g}^2 D_0 \left( {m_{\tilde d_s } ,m_{\tilde d_t } ,m_{\tilde g} ,m_{\tilde g} } \right)} Q_3 \right] \\ 
  \;\;\;\left. \;\;\;\;\;\;\,\; 
 + V_{s\;23}^{d\;LR} V_{t\;23}^{d\;RL} 
  \left[ { - 132 D_2 \left( {m_{\tilde d_s } ,m_{\tilde d_t } , 
        m_{\tilde g} ,m_{\tilde g} } \right) Q_4 
  - 180 D_2 \left( {m_{\tilde d_s } ,m_{\tilde d_t } ,
        m_{\tilde g} ,m_{\tilde g} } \right)} Q_5 \right] \right] .\\ 
 \end{array} \\
 \end{array}
 \label{Heff}
\end{equation}
For definiteness we have quoted \eq{Heff} for $B_s$ mixing but the translation
to other processes is trivial. The definitions of the operators can be found
in Ref.~\cite{Ciuchini:1998ix,Becirevic:2001,Ciuchini:2006dx}, for the
remaining ingredients we follow Ref.~\cite{Crivellin:2008mq} as usual.  $s$
and $t$ label the squark mass eigenstates and a sum over $s,t$ from 1 to 6 is
understood. In the limit of two mass insertions and degenerate squark masses
equation \eq{Heff} simplifies to the result of \cite{Becirevic:2001} by
substituting
\begin{equation}
\renewcommand{\arraystretch}{1.4}
\begin{array}{l}
 D_{0,2} \left( {m_{\tilde q_s } ,m_{\tilde q_t } ,m_{\tilde g} ,m_{\tilde g}
 } \right) \; \to \; 
 F_{0,2} \left( {m_{\tilde q} ,m_{\tilde q} ,m_{\tilde q} ,m_{\tilde q} ,m_{\tilde g} ,m_{\tilde g} } \right) \\[1mm] 
 V^{q\,LL}_{s\,fi}   \to m_{\tilde q} ^2 \delta^{d\,LL} _{fi} ,\,\,
 \quad 
 V^{q\,RR}_{s\,fi} \to m_{\tilde q} ^2 \delta^{q\,RR}_{fi},\,\, 
 \quad 
 V^{q\,RL}_{s\,fi} \to m_{\tilde q} ^2 \delta^{q\,RL}_{fi} ,\,\,
 \quad 
 V^{q\,LR}_{s\,fi } \to m_{\tilde q} ^2 \delta^{q\,LR}_{fi} . \\ 
 \end{array}
\end{equation}
{\eq{Heff} agrees with {the result} in
  Ref.~\cite{Baek:2001kh} {and corrects} two color
  factors in Eq.~II.9 of Ref.~\cite{Hagelin:1992}.} In
  Ref.~\cite{Ciuchini:2006dw}, a NLO calculation of the effective
  $\Delta F=2$ Hamiltonian has been carried out.  The authors reduce
  the theoretical uncertainty and find corrections of about 15 percent
  to the LO result. They miss our effects from the flavor-changing
  self-energies, because they work with degenerate squark masses, so
  that the self-energy contributions cancel as discussed at the end of
  Sect.~\ref{sect:2}.  As we will see, including the renormalized
  vertices can yield an effect of 1000\%\ and more. So it is
  sufficient for our purpose to stick to the LO Hamiltonian of
  \eq{Heff} with the renormalized vertices of Sect.~\ref{Heff}.  To
  incorporate the large logarithms from QCD we use the RG evolution
  computed in Refs.~\cite{Becirevic:2001,Buras:2001ra}.  For the bag
  factors parameterizing the hadronic matrix elements we take the
  lattice QCD values of Ref.~\cite{Lubicz:2008am}.  We show the
  effects of the renormalization of the squark-quark-gluino vertex on
  $\Delta M_{d,s}$ and the general pattern of the new contributions in
  the following subsection.

\subsection{\boldmath \bbm\unboldmath}
The amplitude of \bbmq, $q=d$ or $s$, is conventionally denoted by $M_{12}$.
New physics contributions will typically change magnitude and phase of this
amplitude. $|M_{12}|$ is probed through the mass difference $\dm_q$ among the
two mass eigenstates of the \bb\ system, while any new contribution to $\arg
M_{12}$ will modify certain CP asymmetries. For the formalism and
phenomenology of \bbm\ we refer to Ref.~\cite{run2}, an update of the SM
contributions to \bbm\ can be found in Ref.~\cite{ln2}.

The chirally enhanced contributions are important for the constraints on
$\delta^{d\,LR}_{ij}=\delta^{d\,RL\,*}_{ji}$. They are also relevant if one
seeks constraints on $\delta^{d\,LL}_{ij}$ in the
large-$\tan\beta$ region, but for this case $\rm{Br}[B\to\rm{X_s}\gamma] $ is
more powerful.  Therefore we restrict our discussion in this section to the LR
elements, for which our new effects lead to drastic changes in the
SUSY-contributions to $\dm_q$. We denote the result with renormalized
squark-quark-gluino vertices by $\dm_{q,\rm ren}$.  In
\fig{B-Mischung-Verhalten} we show the ratio of $\dm_{q,\rm ren}$ to the LO
result $\dm_{q,\rm LO}$, which is calculated from the gluino-squark box
diagram without our new contributions.  As one can easily see, the effects
from the finite vertex renormalization drop out for degenerate squark masses,
while dramatic effects for large and unequal squark masses are observed: For
instance, a value of $\dm_{q,\rm ren}/\dm_{q,\rm LO} =50$ implies that the
constraint on the studied $\delta^{d\,AB}_{ij}$ is stronger by a factor of
$\sqrt{50}\approx 7$, because both $\dm_{q,\rm ren}$ and $\dm_{q,\rm LO}$ are
practically quadratic in $\delta^{d\,AB}_{ij}$ (cf.~MIA to see this).
\begin{nfigure}{t}
  \includegraphics[width=0.48\textwidth]{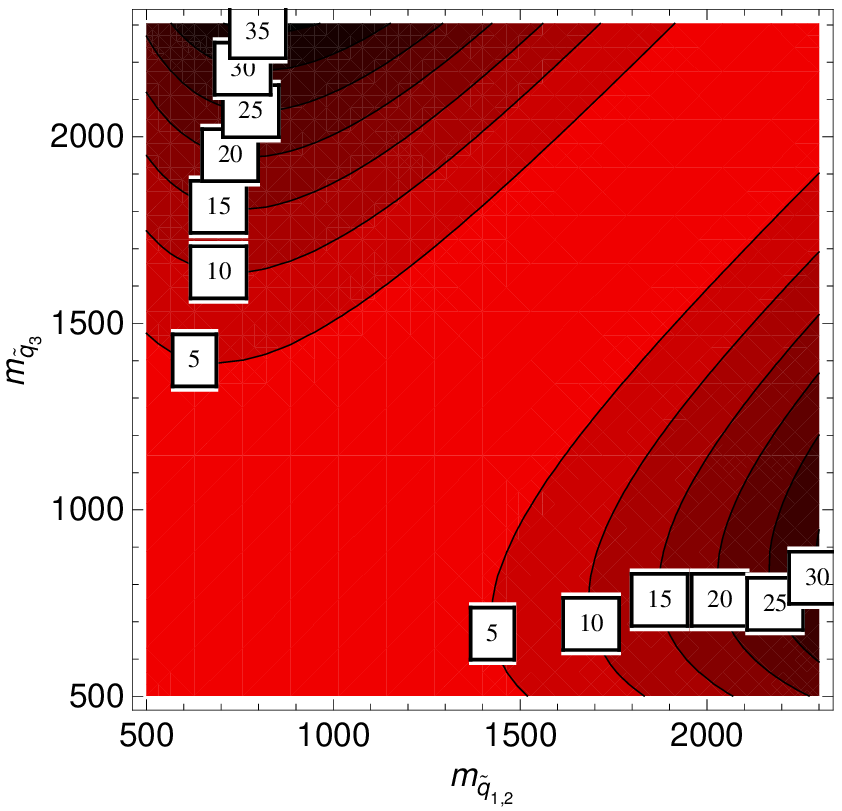} ~~~
  \includegraphics[width=0.48\textwidth]{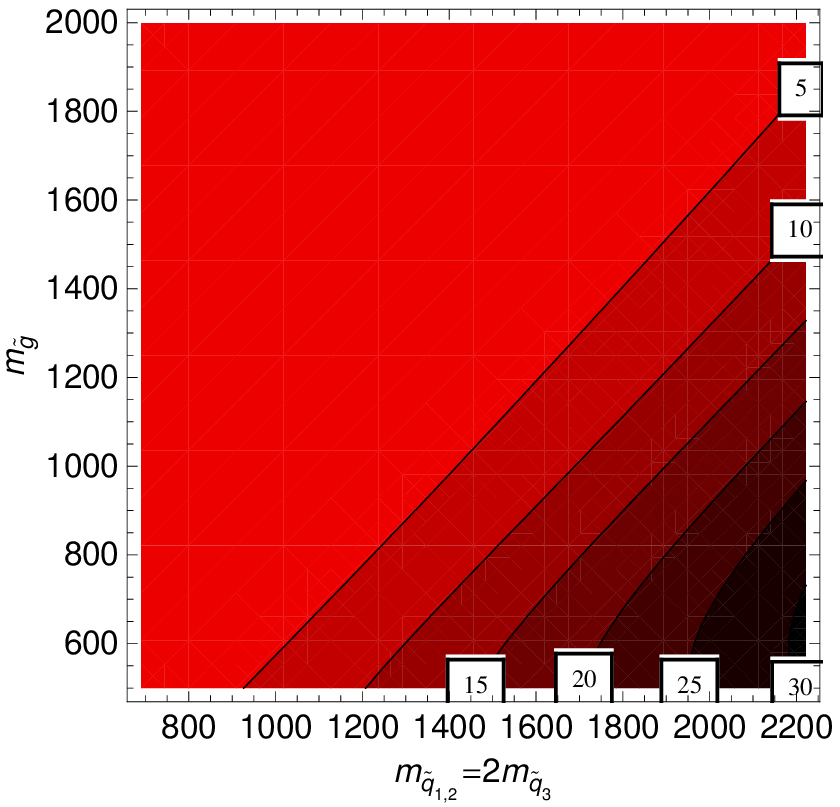}
\caption{Left: Contour plot of 
  $\dm_{q,\rm ren}/\dm_{q,\rm LO}$ as a function of
  $m_{\tilde{q}1,\tilde{q}2}$ and $m_{\tilde{q}3}$ with
  $m_{\tilde{g}}=1000\,\textrm{GeV}$. Right: $\dm_{q,\rm ren}/{\Delta
    M_B}_{\rm LO}$ as a function of
  $m_{\tilde{q}1,\tilde{q}2}=2m_{\tilde{q}3}$ and $m_{\tilde{g}}$.  The
  numbers in the little squares denote the value of $\dm_{q,\rm
  ren}/\dm_{q,\rm LO}$ of the corresponding contour. 
\label{B-Mischung-Verhalten}}
\end{nfigure}
We remark that the value of $\tan\beta$ is inessential in this section,
varying $\tan\beta$ leads to ${\cal O}(1\%)$ changes of our $\Delta F=2$
results.

In order to determine the possible size of new physics (NP), it is necessary
to know the SM contribution to the process in question.  For \bbm, this first
requires the control over hadronic uncertainties, which presently obscure the
quantification of NP contributions from the precise data on $\dm_d$
\cite{Amsler:2008zzb} and $\dm_s$ \cite{Abazov:2007nw,Abulencia:2006ze}. In
the case of \bbmd\ one must also address $V_{td}$, because \bbmd\ is used to
determine this CKM element through the usual fit to the unitarity triangle. A
first analysis combining different quantities probing the \bbms\ amplitude has
revealed a $2\sigma$ discrepancy of $\arg M_{12}$ with the SM prediction
\cite{ln2}.  Since then the CKMfitter \cite{CKMfitter} and UTfit
collaborations \cite{Ciuchini:2000de} have constrained the possible
contributions of new physics to the \bbmd\ and \bbms\ amplitudes with
sophisticated statistical (Frequentist and Bayesian, respectively) methods,
using new information on $\arg M_{12}$ gained from tagged $B_s\to J/\psi
\phi$ data \cite{taggedphaseCDF,taggedphaseD0}.  
We use the corresponding recent UTfit analysis as shown
in \fig{UTfit-B-mixing}.
\begin{nfigure}{tb}
\includegraphics[width=0.48\textwidth]{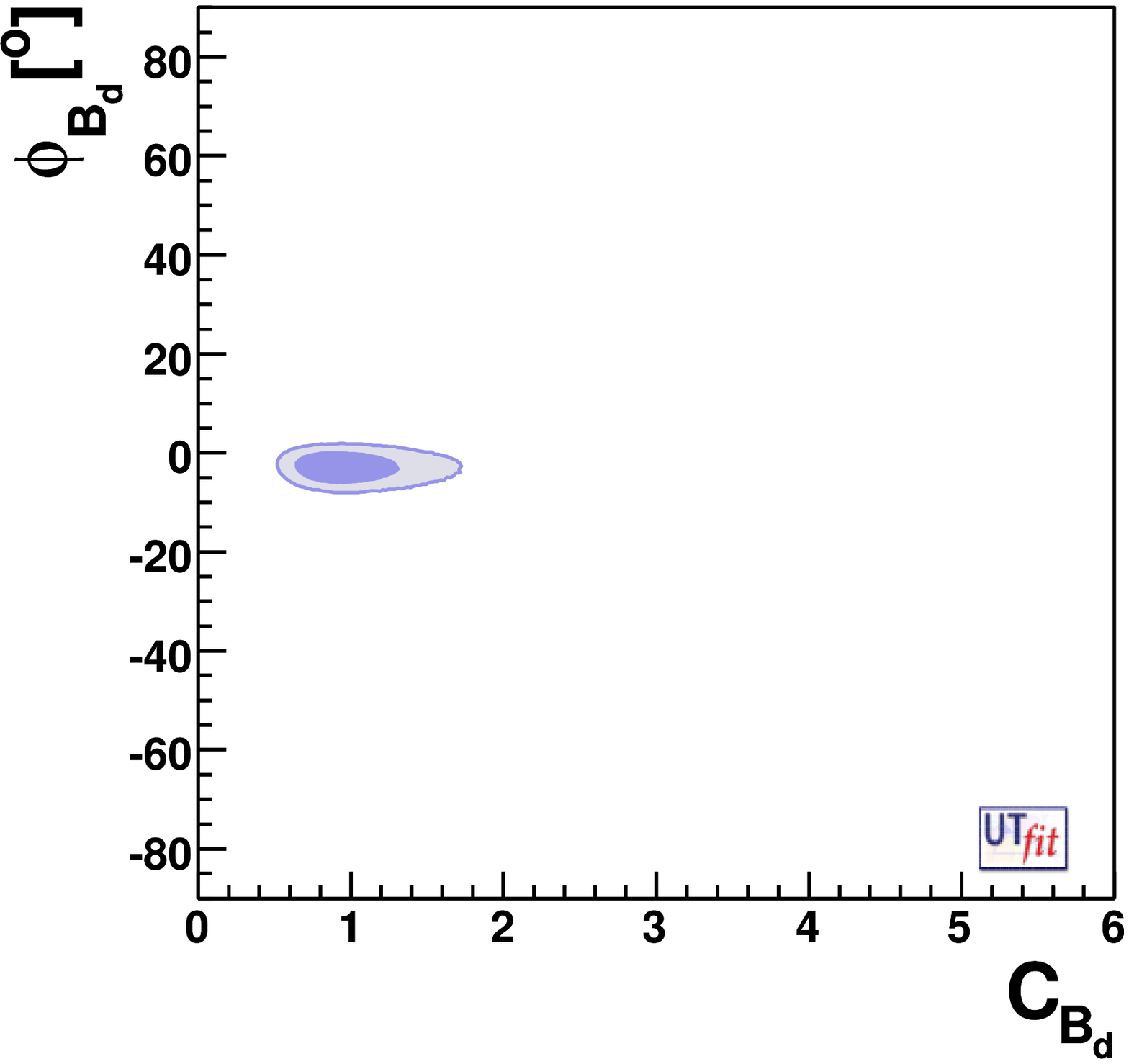}
~~~
\includegraphics[width=0.48\textwidth]{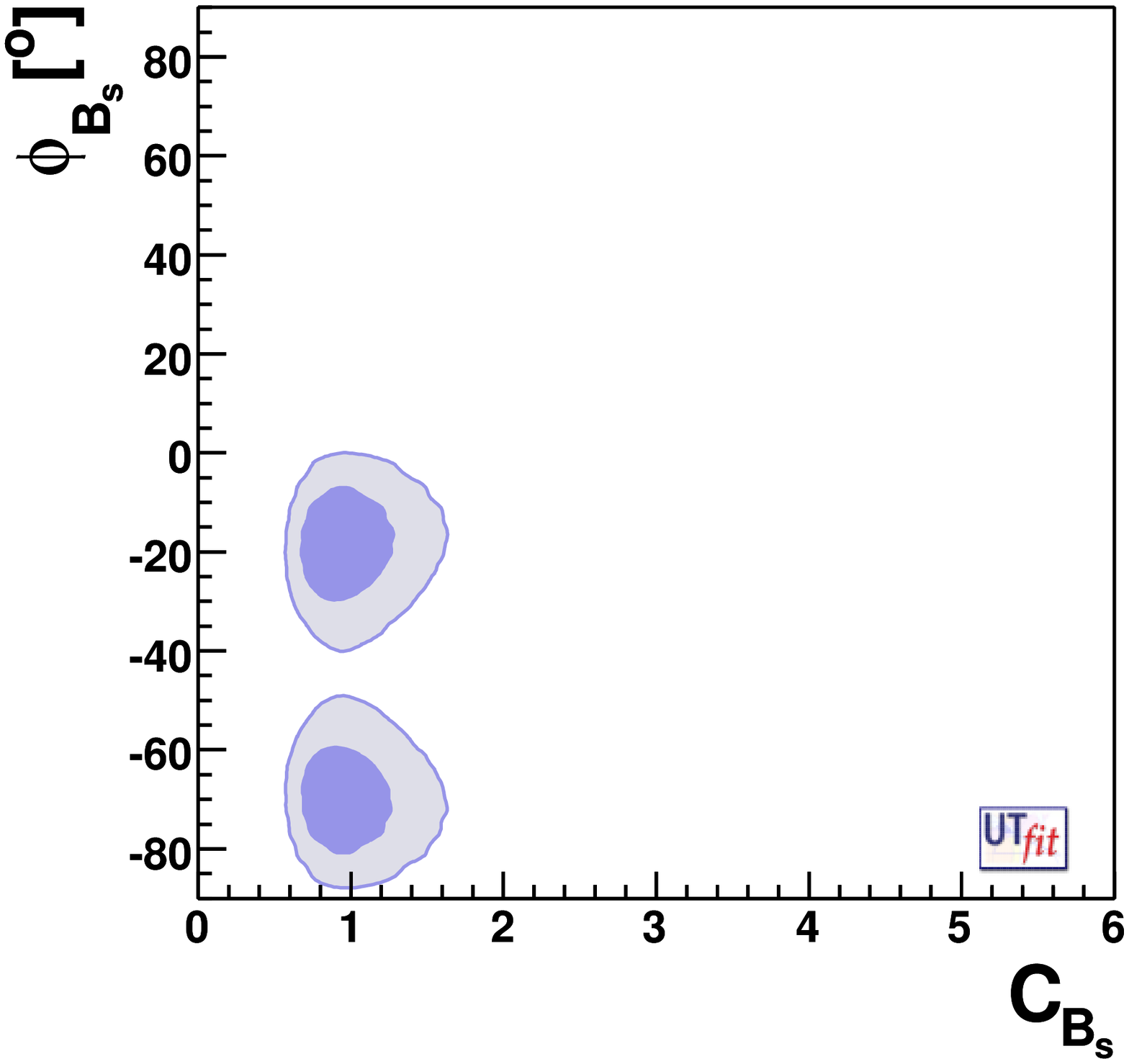}
\caption{
  Allowed range for NP-contributions to \bbmq, $q=d,s$, in the
  $\phi_{B_q}-C_{B_{q}}$ plane taken from the web update of 
  Ref.~\cite{Bona:2008jn}.
  See \eq{NP-B-mixing} and related text for details. 
  For a related CKMfitter analysis see Ref.~\cite{tiss}.
  \label{UTfit-B-mixing}}
\end{nfigure}
The quantities $C_{B_q}$ and $\phi_{B_q}$ shown in the plots are defined as:
\begin{equation}
C_{B_q} e^{2 i \phi _{B_q } } = \Delta_q 
  = \frac{\bra{B_q} H_{\rm eff} \ket{\Bbar_q}}{
          \bra{B_q} H_{\rm eff}^{\rm SM} \ket{\Bbar_q}}
  = \frac{M_{12}}{M_{12}^{\rm SM}}
  = \frac{|M_{12}^{\rm SM}| + 
          |M_{12}^{\rm NP}| e^{ 2 i \phi _{\rm NP} }}{
          |M_{12}^{\rm SM}| }
 \label{NP-B-mixing}
\end{equation}
Here $\phi _{\rm NP}$ is the difference between the phase of the new physics
contribution $M_{12}^{\rm NP}$ and the phase of the SM box diagrams.
Refs.~\cite{ln2} and \cite{tiss} show the experimental constraints in the
complex $\Delta_q$ planes instead.  The plots in \fig{B-Mischung-komplex} show
the allowed regions in the complex $\delta^{d\,LR}_{23}$ and
$\delta^{d\,LR}_{13}$ planes.  The analogous constraints on the complex
$\delta^{d\,RL}_{23}$ and $\delta^{d\,RL}_{13}$ planes look identical, because
$|\Delta F|=2$ processes are parity-invariant. To obtain
\fig{B-Mischung-komplex} we have parameterized the border of the 95\%\ CL
region in \fig{UTfit-B-mixing} and determined the values
$\rm{Re}[\delta^{d\,AB}_{13,23}]$ and $\rm{Im}[\delta^{d\,AB}_{13,23}]$ which
correspond to this region by using \eq{NP-B-mixing} with $M_{12}^{\rm NP}$
calculated from $H_{\rm eff}^{\rm SUSY}$ in \eq{Heff}. The hadronic matrix
elements are conventionally expressed in terms of the product of the squared
decay constant $f_{B_q}^2$ and a bag factor. The dependence on $f_{B_q}$ drops
out in the ratio defining $C_{B_q} e^{2 i \phi _{B_q } }$, which only involves
the ratios of the bag factors of the different operators.  That is, the
sizable uncertainty of $f_{B_q}$ does not enter at this step, but entirely
resides in the allowed region for $C_{B_q} e^{2 i \phi _{B_q } }$ plotted in
\fig{UTfit-B-mixing}. Therefore our results in \fig{B-Mischung-komplex}
correspond to the ranges for $f_{B_q}\sqrt{B_q}$ (where $B_q$ is the bag
factor of the SM operator) used in (the web update of)
Ref.~\cite{Bona:2008jn}. These ranges are $f_{B_s}\sqrt{B_s}=(270\pm
30)\,\textrm{MeV}$ and $f_{B_s}\sqrt{B_s}/(f_{B_d}\sqrt{B_d})= 1.21\pm 0.04$
(both at 1$\sigma$).
\begin{nfigure}{tb}
\includegraphics[width=0.48\textwidth]{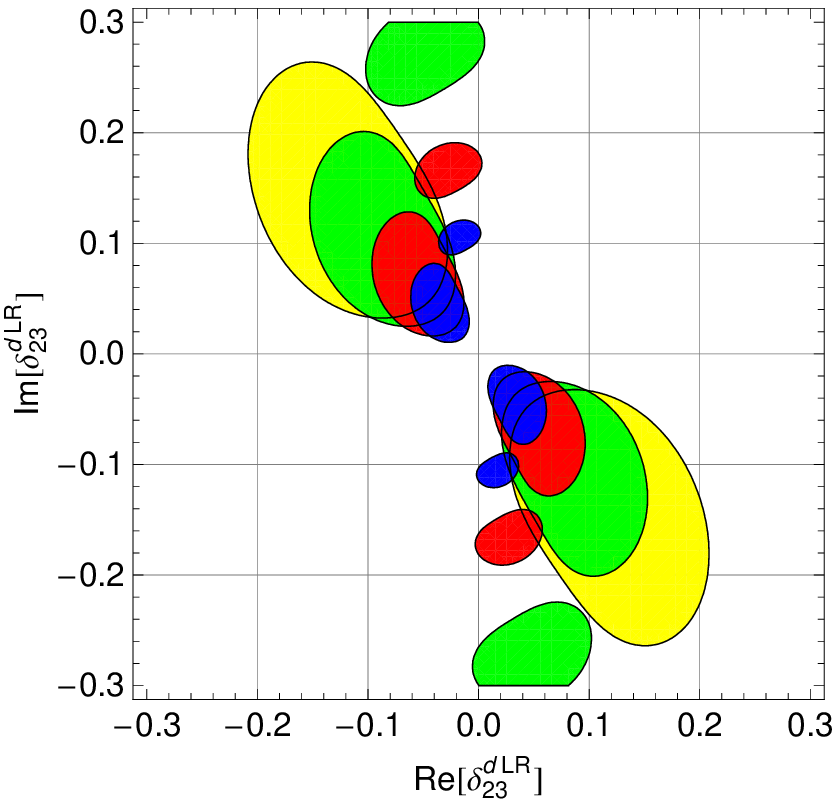}
~~~
\includegraphics[width=0.48\textwidth]{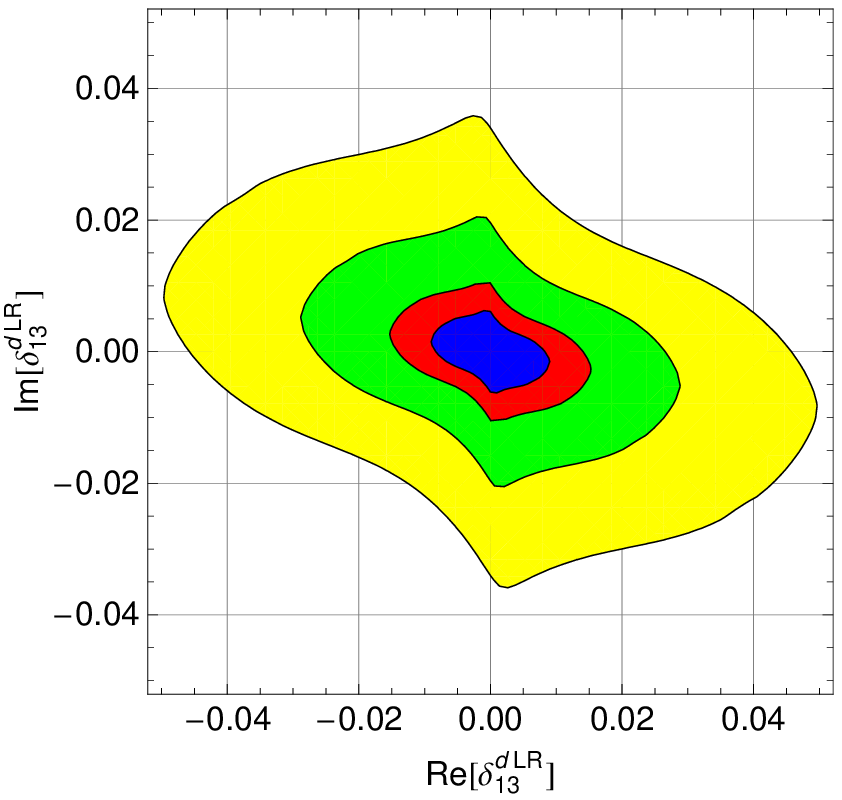}
\caption{
  Allowed regions in the complex $\delta^{d\,LR}_{13,23}$-plane from
  $B_s$-mixing (left plot) and $B_d$-mixing (right plot) with $m_{\tilde
    {q}1,2}=2m_{\tilde q3}=1000\,\textrm{GeV}$. The yellow, green, red and
  blue (light to dark) areas  correspond to the 95\%\ CL regions for
  $m_{\tilde{g}}=1200\,\textrm{GeV}$, $m_{\tilde{g}}=1000\,\textrm{GeV}$,
  $m_{\tilde{g}}=800\,\textrm{GeV}$, $m_{\tilde{g}}=600\,\textrm{GeV}$. The
  effect is practically insensitive to $\tan\beta$.  The same constraints are
  obtained for the complex $\delta^{d\,RL}_{13,23}$-plane.
  \label{B-Mischung-komplex}}
\end{nfigure}
In the case of \bbms\ the colored regions corresponding to different gluino
masses do not overlap and do not contain the point $\delta^{d\,LR}_{23}=0$,
because the SM value $M_{12}^{\rm SM}$ is not in the 95\%\ CL region of the
UTfit analysis.  The tension with the SM largely originates from the $B_s\to
J/\psi \phi$ data \cite{taggedphaseCDF,taggedphaseD0}. A large gluino mass
suppresses the supersymmetric contribution to $M_{12}$, so that a larger value
of $\mbox{Im}\, \delta^{d\,LR}_{23}$ is needed to bring $\arg M_{12}$ into
the 95\%\ CL region.

\subsection{\boldmath \kkm\unboldmath}
In \kkm\ the situation is very different from \bbm, because the chiral
enhancement factor $A^d_{12}/m_{s}$ involves the small $m_s$ rather than
$m_b$. The observed smallness of FCNC transitions among the first two
generations not only forbids large $\delta^{q\,AB}_{12}$ elements but also
constrains the splittings among the squark masses of the first two generations
severely.  This observation suggests the presence of a $U(2)$ symmetry
governing the flavor structure of the first two (s)quark generations. This
symmetry cannot be exact, as it is at least broken by the difference
$Y^{s}-Y^{d}$ of Yukawa couplings. That is, the numerical size of
flavor-$U(2)$ breaking is somewhere between $10^{-4}$ and a few times
$10^{-2}$, depending on the size of $\tan\beta$. We may therefore fathom
deviations from flavor universality in the same ball-park in the squark sector.
Counting $\Sigma _{12}^{d\,RL}$ as first-order in some $U(2)$-breaking
parameter, we realize that our chiral enhancement factors are of zeroth order
in $U(2)$ breaking due to the appearance of the factor $1/m_s$ in \eq{DeltaU}.
Therefore \kkm\ is extremely sensitive to the remaining source of
flavor-$U(2)$ breaking in the problem, the mass splitting $m_{\tilde q2}-
m_{\tilde q1}$. At this point we mention that it is important to control the
renormalization of $m_s$ in the presence of ordinary QCD corrections. In
Appendix~B of Ref.~\cite{Hofer:2009xb} it has been shown that all QCD
corrections combine in such a way that the inverse power of $m_s$ is the
$\overline{\rm MS}$ mass evaluated at the scale $Q=M_{\rm SUSY}$, provided
that gluonic QCD corrections to $\Sigma _{12}^{d\,RL}$ are also calculated in
the $\overline{\rm MS}$ scheme.

The sensitivity of the chirally enhanced corrections to the squark-mass
splitting is displayed in \fig{K-Mischung-Verhalten}. Constraints on
$\delta^{d\,AB}_{12}$ from \kkm\ have been considered for a long time (see
Refs.~\cite{Gabbiani:1996,Ciuchini:1998ix}).  Again we use the UTfit analysis
(cf.\ the left plot of \fig{UTfit-K-mixing}) exploiting the mass difference
$\dm_K$ and the CP-violating quantity $\epsilon_K$. We show our improved
constraints on the complex $\delta^{d\,LR}_{12}$ element in the right plot of
\fig{UTfit-K-mixing}.
\begin{nfigure}{tb}
\includegraphics[width=0.48\textwidth]{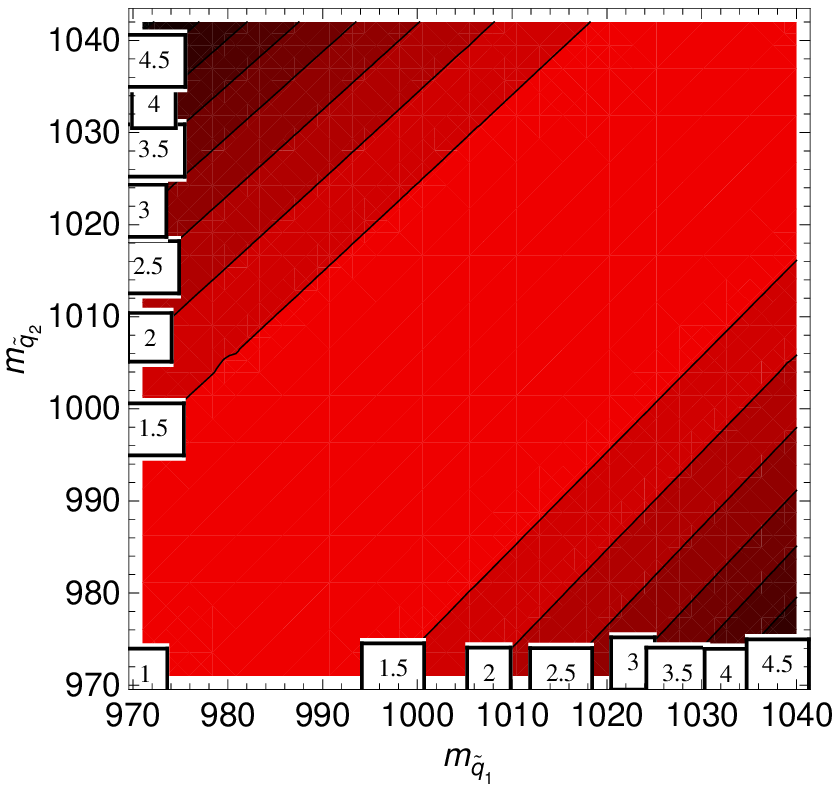}
~~~
\includegraphics[width=0.48\textwidth]{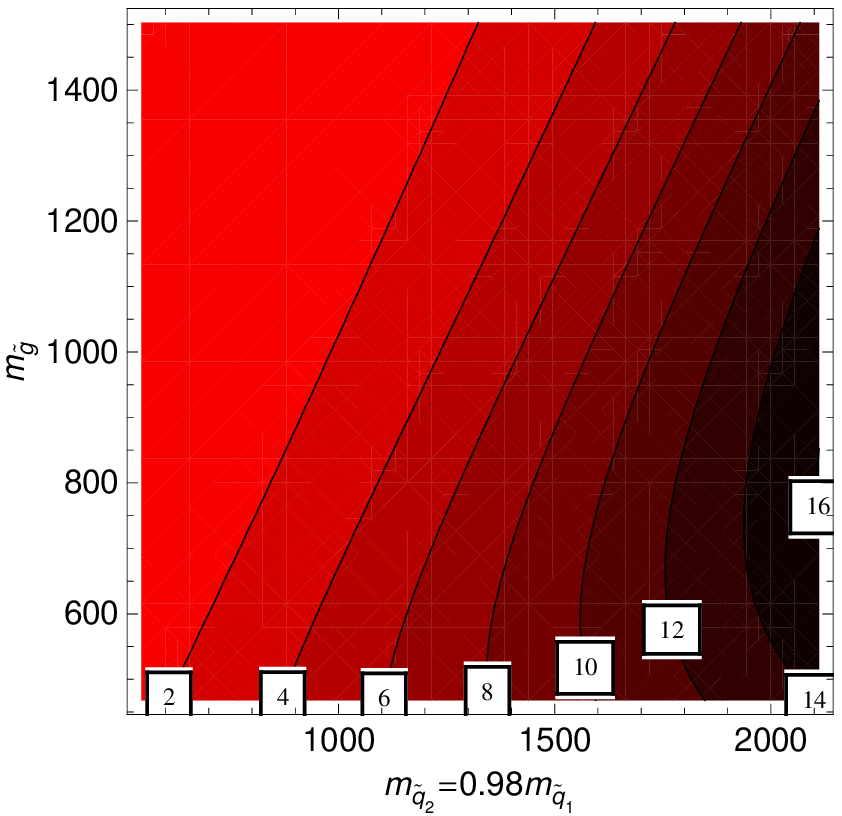}
\caption{
Left: Contour plot of ${\dm_K}_{\rm ren}/{\dm_K}_{\rm LO}$ as a function of 
$m_{\tilde{q}1}$ and $m_{\tilde{q}2}$ with $m_{\tilde{g}}=1000\,\textrm{GeV}$.
Right: ${\Delta M_K}_{\rm ren}/{\Delta M_K}_{\rm LO}$ as a function of 
$m_{\tilde{q}2}=0.98m_{\tilde{q}1}$ and $m_{\tilde{g}}$.
\label{K-Mischung-Verhalten}}
\end{nfigure}
\begin{nfigure}{tb}
\includegraphics[width=0.48\textwidth]{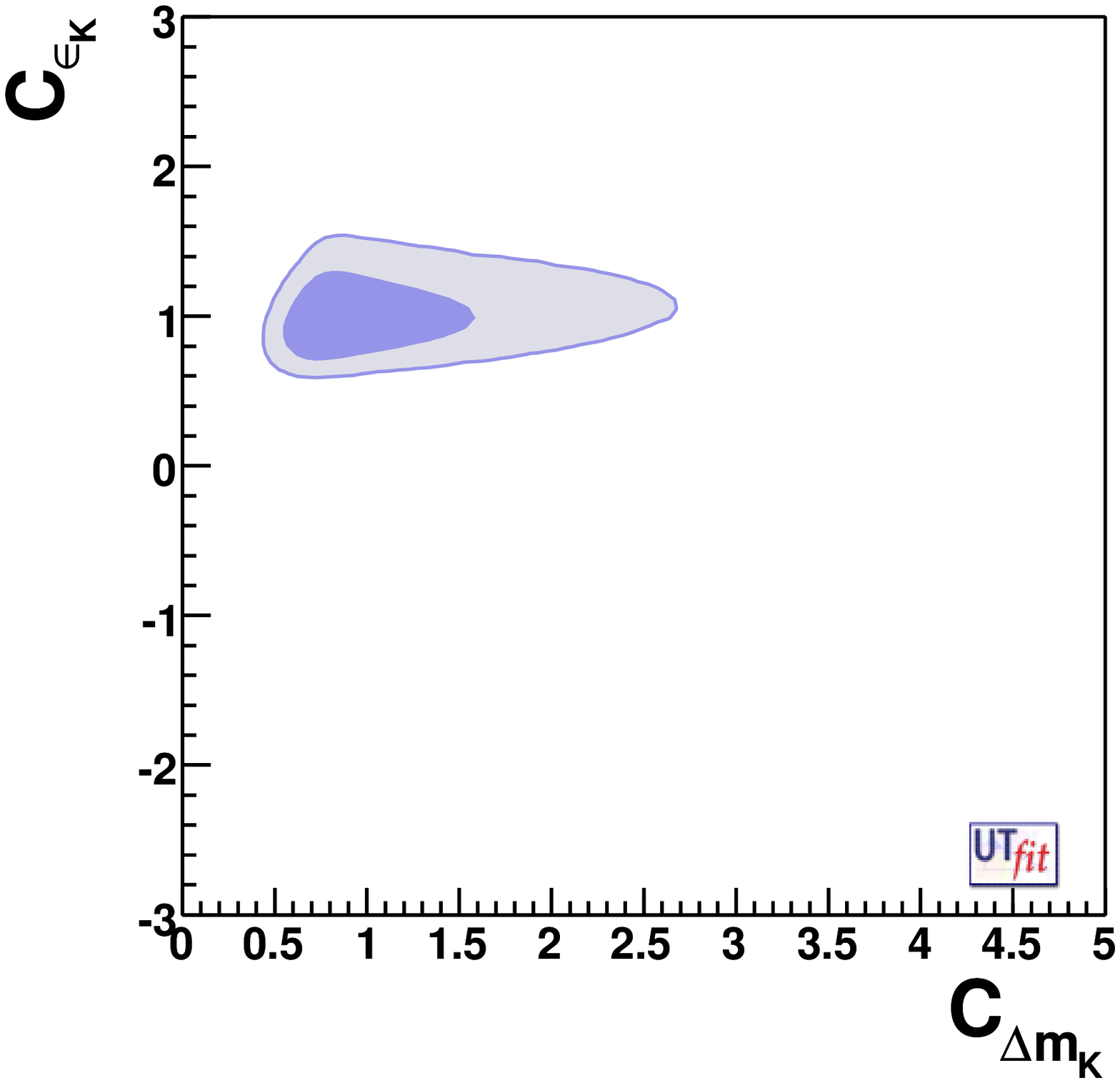}
\includegraphics[width=0.48\textwidth]{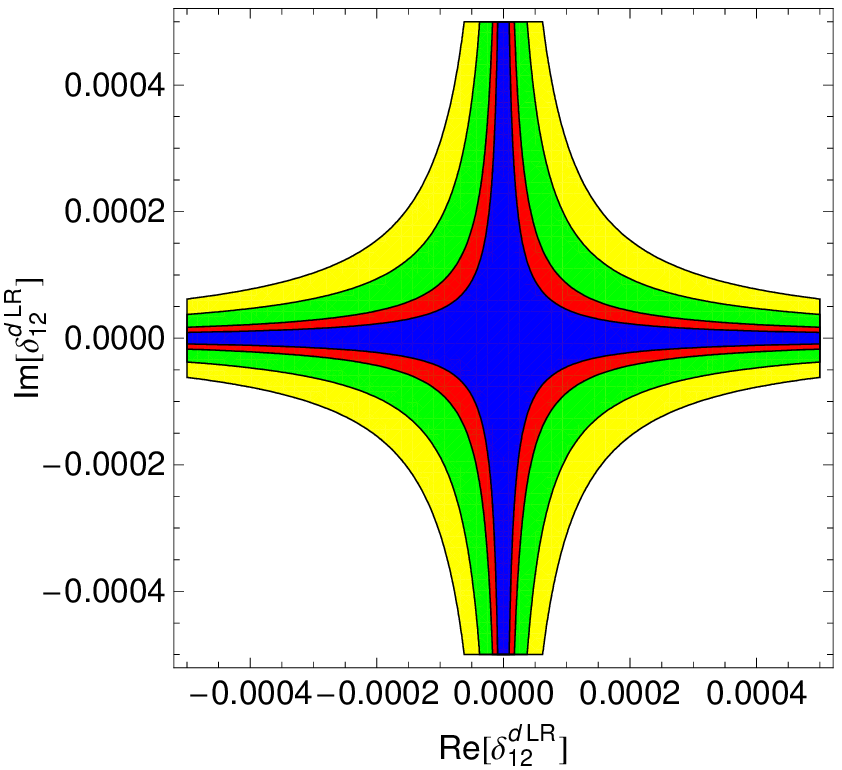}
\caption{Left: Allowed range for NP contributions to K mixing determined 
  by the UTFit collaboration 
  \cite{Ciuchini:2000de}. The light region corresponds to $95\%$ CL.
  Right: Allowed region in the complex
  $\delta^{d\,LR}_{12}$-plane with
  $m_{\tilde{g}}=750\,\textrm{GeV}$, $m_{\tilde{q}1}=1000\,\textrm{GeV}$.
  Yellow: $m_{\tilde{q}2}=1000,\textrm{GeV}$, green:
  $m_{\tilde{q}2}=999\,\textrm{GeV}$, red: $m_{\tilde{q}2}=998\,\textrm{GeV}$,
  blue: $m_{\tilde{q}2}=997\,\textrm{GeV}$ (light to dark). 
  The same constraints are found for $\delta^{d\,RL}_{12}$.
  \label{UTfit-K-mixing}}
\end{nfigure}
Figs.~\ref{K-Mischung-Verhalten} and \ref{UTfit-K-mixing} are analogous to
Figs.~\ref{B-Mischung-Verhalten}--\ref{B-Mischung-komplex} addressing \bbm; we
refer to the corresponding figure captions for further explanation. We find
that \kkm\ indeed probes flavor violation in the squark sector of the first
two generations at the per-mille level. The constraints on
$\delta^{d\,LR}_{12}$ sharply grow with $|m_{\tilde q2}- m_{\tilde q1}|$.

\section{Conclusions}
In the generic MSSM FCNC processes proceed through gluino-squark loop
diagrams. In the present paper we have pointed out that these processes
receive chirally enhanced corrections which contribute to the FCNC processes
via two-loop or three-loop diagrams but can numerically dominate over the
usual LO one-loop diagrams. The chirally enhanced contributions involve a
flavor change in a self-energy sub-diagram attached to an external leg of the
diagram. These effects can be absorbed into a finite renormalization of the
squark-quark-gluino vertex. Our new effects vanish if the squark masses are
degenerate. Their relative importance with respect to the LO diagrams is
further larger for heavier squarks. We have consistently diagonalized 
the squark mass matrices exactly, thereby avoiding the mass insertion
approximation. In this way the left-right mixing of bottom squarks 
is included correctly and the large-$\tan\beta$ region can be accessed.  

Our first phenomenological study addresses $B\to X_s \gamma$. In this
process our new effects are only relevant if $|\mu|\tan\beta $ is
large.  Taking $\tan\beta=50$ we present new bounds on the four
quantities $\delta^{d\,LL}_{23}$, $\delta^{d\,LR}_{23}$,
$\delta^{d\,RL}_{23}$, and $\delta^{d\,RR}_{23}$ which parameterize
the off-diagonal elements of the down-squark mass matrix linking
strange and bottom squarks.  These bounds are depicted in
\fig{b-s-gamma-constraints} for the case of real MSSM parameters.  As
a general pattern we find that the chirally enhanced contributions
decrease the size of the SUSY contribution to $\rm{Br}[B\to {\rm
X_s}\gamma]$ if $\mu$ is positive. Conversely, the chirally enhanced
two-loop contributions increase the SUSY contribution to
$\delta^{d\,AB}_{23}$ for $\mu<0$.  That is, for positive values of
$\mu$, which are preferred by the anomalous magnetic moment of the
muon, the bounds on $\delta^{d\,AB}_{23}$ become weaker.
A reduced sensitivity of $\rm{Br}[B\to {\rm X_s}\gamma]$ to $\delta^{d\,AB}_{23}$ is welcome
for models in which the Yukawa couplings of the first two generations
are zero at tree-level and light quark masses and all off-diagonal CKM
elements are generated radiatively from the soft SUSY-breaking terms
\cite{Borzumati:1997bd,Crivellin:2008mq}. In
Ref.~\cite{Crivellin:2008mq} it has been shown that this idea complies
with present-day data on FCNC processes. With the weaker bound on
$\delta^{d\,LR}_{23}$, which is needed to generate $V_{cb}$, therefore
a larger portion of the MSSM parameter space becomes compatible with
the radiative generation of flavor.

The second analysis of this paper is devoted to $B_d$, $B_s$, and $K$
mixing.  Using the data on the mass differences $\dm_d$ and $\dm_s$
and on CP asymmetries we find new constraints on the complex
$\delta^{d\,LR}_{13}$, $\delta^{d\,LR}_{23}$, $\delta^{d\,RL}_{13}$,
and $\delta^{d\,RL}_{23}$ elements (see \fig{B-Mischung-komplex}).  In
most of the parameter space the constraints become much stronger
compared to the LO analysis if the sbottom mass differs sizably from
the squark masses of the first two generations, irrespective of the
size of $\tan\beta$. \kkm\ is
even more sensitive to the chirally enhanced self-energies, provided
there is a non-zero mass splitting among the squarks of the first two
generations.  As illustrated in \fig{UTfit-K-mixing} already mass
splittings in the sub-percent range strengthen the bounds on
$\delta^{d\,LR}_{12}$ and $\delta^{d\,RL}_{12}$ severely.

After completion of the presented work, a complete NLO calculation of
supersymmetric QCD corrections to $\Delta F=2$ Wilson coefficients with exact
diagonalization of the squark mass matrices and non-degenerate squark masses
has appeared \cite{virto}. However, our chirally enhanced effects are not
included, because Ref.~\cite{virto} adopts the definition of the super-CKM
basis based on an on-shell definition of the quark fields, as described in our
Sect.~2 after \eq{DeltaU}). With this definition the chirally enhanced effects
are implicitly contained in the numerical values of the $\delta^{q\,AB}_{ij}$.
While the NLO corrections of Ref.~\cite{virto} are typically numerically much
smaller than ours, they are important to control the scale and scheme
dependences of the LO diagrams.  Therefore our results in Sect.~4 and those of
Ref.~\cite{virto} are complementary to each other.

\section*{Acknowledgments} 
The authors thank Lars Hofer and Dominik Scherer for many fruitful
discussions and Javier Virto for e-mail exchange about the
renormalization of the squark-quark-gluino vertex. {We
further thank the referee for drawing our attention to
Ref.~\cite{Foster}.}  This work is supported by BMBF under grant
no.~05H09VKF and by the EU Contract no.~MRTN-CT-2006-035482, \lq\lq
FLAVIAnet''.  Andreas Crivellin acknowledges the financial support of
the State of Baden-W\"urttemberg through \emph{Strukturiertes
Promotionskolleg \lq\lq Elementarteilchenphysik und
Astroteilchenphysik''}.

\end{document}